\documentclass[twocolumn,letterpaper]{IEEEAerospaceCLS}

\usepackage[utf8]{inputenc}
\usepackage[]{graphicx}
\usepackage{amsmath}
\usepackage{amsfonts}
\usepackage[version=4]{mhchem}
\usepackage{siunitx}
\usepackage{listings}
\usepackage{longtable}
\usepackage{tabularx}
\usepackage[table]{xcolor}
\usepackage{caption}
\usepackage{subcaption}
\usepackage{hyperref}
\usepackage{kantlipsum,cuted}
\usepackage{dblfloatfix}

\usepackage{babel}
\usepackage{inputenc}
\usepackage{tcolorbox}
\usepackage{xcolor}
\definecolor{txtgreen}{RGB}{50,156,40}
\definecolor{txtorange}{RGB}{189,101,34}
\definecolor{txtpurple}{RGB}{212,17,208}
\definecolor{txtblue}{RGB}{50,76,168}

\usepackage{algorithm}
\usepackage{algpseudocode}

\newcommand\norm[1]{\left\lVert#1\right\rVert}

\newcommand{\ignore}[1]{} 

\begin{document}
\title{\Large \bf Precise Distributed Satellite Navigation: Differential GPS with Sensor-Coupling for Integer Ambiguity Resolution}

\author{
    Samuel Y. W. Low \\ 
    Stanford University, Stanford, CA 94305 \\
    \href{sammmlow@stanford.edu}{sammmlow@stanford.edu}
    \and 
    Simone D'Amico \\
    Stanford University, Stanford, CA 94305 \\
    \href{damicos@stanford.edu}{damicos@stanford.edu}
    \thanks{2024 IEEE AERO @ BIG SKY, MONTANA (PRE-PRINT)}
}

\maketitle

\thispagestyle{plain}
\pagestyle{plain}

\maketitle

\thispagestyle{plain}
\pagestyle{plain}


\begin{abstract}
Precise relative navigation is a critical enabler for distributed satellites to achieve new mission objectives impossible for a monolithic spacecraft. Carrier phase differential GPS (CDGPS) with integer ambiguity resolution (IAR) is a promising means of achieving cm-level accuracy for high-precision Rendezvous, Proximity-Operations and Docking (RPOD), In-Space Servicing, Assembly and Manufacturing (ISAM) as well as satellite formation flying and swarming. However, IAR is extremely sensitive to received GPS signal noise, and may fail in adverse environments with severe multi-path or high thermal noise. This paper proposes a sensor-fusion based approach to achieve IAR under such conditions in two coupling stages. A loose coupling stage efficiently fuses through an Extended Kalman Filter the CDGPS measurements with on-board sensor measurements such as range from inter-satellite cross-links, and vision-based bearing angles from a monocular camera. A second tight-coupling stage augments the cost function of the integer weighted least-squares minimization with a soft constraint function using noise-weighted observed-minus-computed residuals from these external sensor measurements. Integer acceptance tests are empirically modified by a coefficient reflecting the added constraints. Partial ambiguity resolution is applied to graduate integer fixing, where a subset of ambiguities that maximizes the probability of success is selected for fixing rather than the full batch of ambiguities. These proposed techniques are packaged into flight-capable software, with ground truths simulated by the Stanford Space Rendezvous Laboratory's $\mathbf{\mathcal{S}}^3$ library using state-of-the-art force modelling with relevant sources of errors, and validated in two scenarios: (1) a high multi-path scenario involving rendezvous and docking in low Earth orbit, and (2) a high thermal noise scenario relying only on GPS side-lobe signals during proximity operations in geostationary orbit. This study demonstrates successful IAR in both cases, using the proposed sensor-fusion approach, thus demonstrating potential for high-precision state estimation under adverse signal-to-noise conditions. 
\end{abstract} 


\tableofcontents

\newpage


\section{Introduction and Background}
\label{section1}

Precise relative navigation is a key enabler of new distributed spacecraft mission concepts, paving the way to overcome fundamental limitations of monolithic spacecraft. Carrier phase differential GPS (CDGPS) with integer ambiguity resolution (IAR) is a promising means of achieving cm-level navigation accuracy in Low Earth Orbit (LEO) and beyond \cite{gpstextbook2006}. CDGPS exploits the error-cancelling effects of differencing carrier phase measurements between two receiving spacecraft. IAR addresses the estimation of the ambiguous number of differenced carrier wave cycles as a float before resolving them into their true \textit{integer} form. After differencing, the IAR step is key to achieving precise navigation since knowledge of the number of differential wave cycles leaves us with a phase measurement where correlated errors have been cancelled and the remnant thermal noise remains at the mm-level. Once integers are successfully fixed, the navigator possesses a precise baseline that can be exploited for high precision estimation of other states or environmental parameters of interest. This makes CDGPS with online IAR a very attractive algorithmic choice for precise real-time state estimation.

CDGPS with IAR is fairly established for applications in low Earth orbit (LEO) via post-facto processing. Examples include the GRACE formation flying mission used for high precision gravimetry \cite{kroes2005grace} and to the TerraSAR-X / TanDEM-X formation for radar interferometry \cite{montenbruck2011tsx}. The first online, real-time implementation of CDGPS without IAR aboard Small Satellites was the PRISMA formation flying demonstration mission \cite{damico2010thesis} \cite{damico2013prisma} in 2010. Float ambiguities were estimated using a low-cost single frequency receiver, achieving 5cm and 1mm/s precision (3D root-mean-square) in \textit{real-time} \cite{damico2012safe}. IAR was not executed due to computational constraints. Since then, the Stanford Space Rendezvous Laboratory \href{https://slab.stanford.edu/}{(SLAB)} has developed the Distributed Multi-GNSS Timing and Localization (DiGiTaL) flight software package capable of real-time IAR and thus precise relative navigation \cite{giralo2019digital} \cite{giralo2021digital} in a Kalman filtering framework. DiGiTaL was initially funded by the NASA Small Spacecraft Development Program \cite{giralo2020dwarf} and has been tailored for the upcoming VISORS distributed telescopy mission (launch due on the SpaceX Transponder-12 in October 2024). SLAB will attempt a first-ever demonstration of real-time in-orbit IAR on-board CubeSats during the VISORS mission, relying solely on GPS L1 signals \cite{visors2023aas}. This work builds on top of the DiGiTaL flight software for VISORS, and extends the applicability of CDGPS with real-time IAR beyond LEO to include harsher signal conditions in-orbit. The high sensitivity of IAR to measurement noise makes achieving IAR challenging in high noise scenarios. This study proposes a two-tiered sensor fusion approach involving a \textit{loose-coupling} stage which performs a joint filter measurement update to include external sensor measurements such as range and bearing angles, and a \textit{tight-coupling} stage where the same external sensor measurements are incorporated into the integer search and optimization step during IAR.

\newpage


\section{The Problem Statement}
\label{section2}

Achieving real-time IAR in environments suffering from adverse noise conditions is challenging for a navigation filter. The covariance of float estimates must converge sufficiently before attempting an integer resolution, typically exceeding a stringent $99\%$ probability of success as the recommended threshold in literature \cite{teunissen1998success}. For GPS L1, $\lambda = 19.05cm$, the measurement noise must be far lower than the maximal quarter-wavelength limit $\lambda/4 \approx 5cm$ so as to resolve wave cycles to their correct integers without ambiguity. An under-confident filter would not achieve a steady-state float ambiguity covariance that satisfies the desired probability of success, while an over-confident filter risks resolving floats into the wrong integer value and severely degrade navigation performance if the wrong fix remains undetected. Therefore, even with additional metrologies provided for sensor fusion, there is a need for careful filter design, tuning, and validation through high-fidelity simulations so as to achieve IAR success under such adverse scenarios. Two such scenarios are investigated in this study: (i) rendezvous and docking near highly reflective structures with multi-path effects in Low Earth Orbit (LEO), and (ii) proximity operations in Geostationary Orbit (GEO) relying only on sidelobe signals with poor $C/N_0$ and thus high thermal noise. In the case of a receiver in LEO, this is compounded by short time-visibility of tracked integers (typically $< 15$ minutes) due to a rapidly changing GPS constellation geometry. These challenging scenarios are explored and elucidated in Section \ref{section6}. This paper seeks to address the challenge of unlocking precise state estimation in these harsh environments through IAR. The problem statement is thus succinctly stated as:

\vspace{3mm}
\begin{tcolorbox}
\begin{center}
    \textbf{How can IAR be achieved between distributed space systems operating under adverse signal thermal noise and multi-path?}
\end{center}
\end{tcolorbox}
\vspace{3mm}

The remainder of this paper is organized as follows: Section \ref{section3} reviews concepts of CDGPS and IAR. Section \ref{section4} reviews relevant literature. Section \ref{section5} details the key contributions of this work and the implementation details of packaging into flight-ready software. Section \ref{section6}  specifies the detailed simulation setup for each scenario. Section \ref{section7} demonstrates and discusses simulated flight results of applying these strategies to challenging environments. Section \ref{section8} concludes by reviewing the contributions made in this work.


\section{Review of CDGPS and IAR}
\label{section3}

This section reviews the preliminaries of CDGPS and IAR in order to facilitate understanding of the state-of-the-art in literature that will be covered in the next section. The standard undifferenced carrier phase measurement model with wavelength $\lambda$, between a receiver's antenna $A$ and a transmitting GPS antenna $P$ as per \cite{gpstextbook2006}, is given by

\begin{equation}
\label{eq1}
\footnotesize
  \lambda \phi_A^{(P)} =
  R_A^{(P)} + \lambda N_A^{(P)} + 
  I + c \left( \delta t_A - \delta t^{(P)} \right)
   + W^{(P)} + \varepsilon_A^{(P)}
\end{equation}

\begin{figure}[ht]
    \centering
    \includegraphics[width=0.45\textwidth]{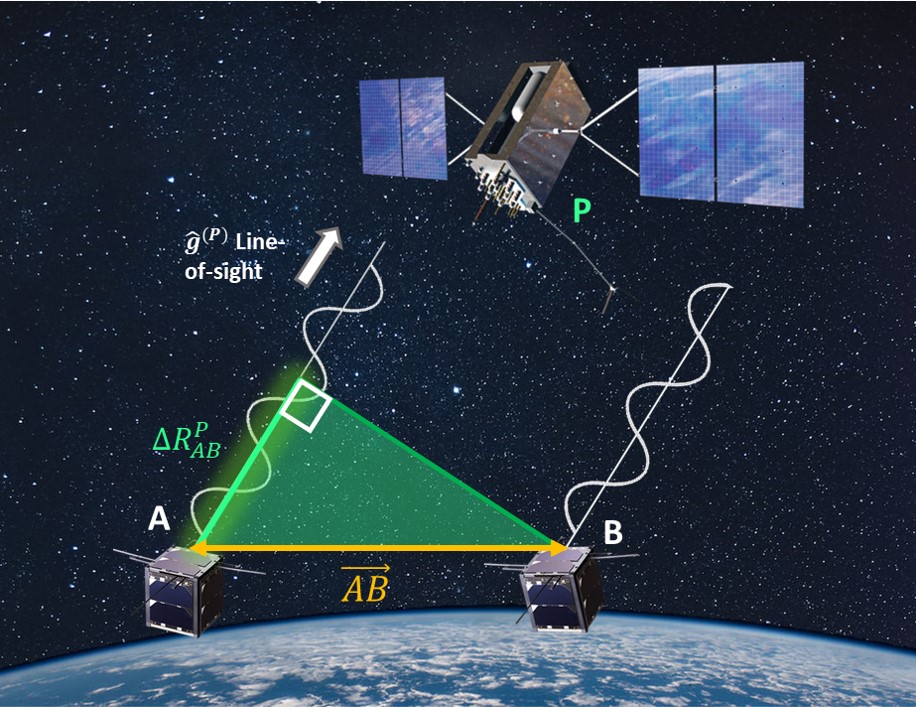}
    \caption{Geometry of single-difference carrier phase}
    \label{figure:review-sdcp}
\end{figure}

\begin{figure}[ht]
    \centering
    \includegraphics[width=0.45\textwidth]{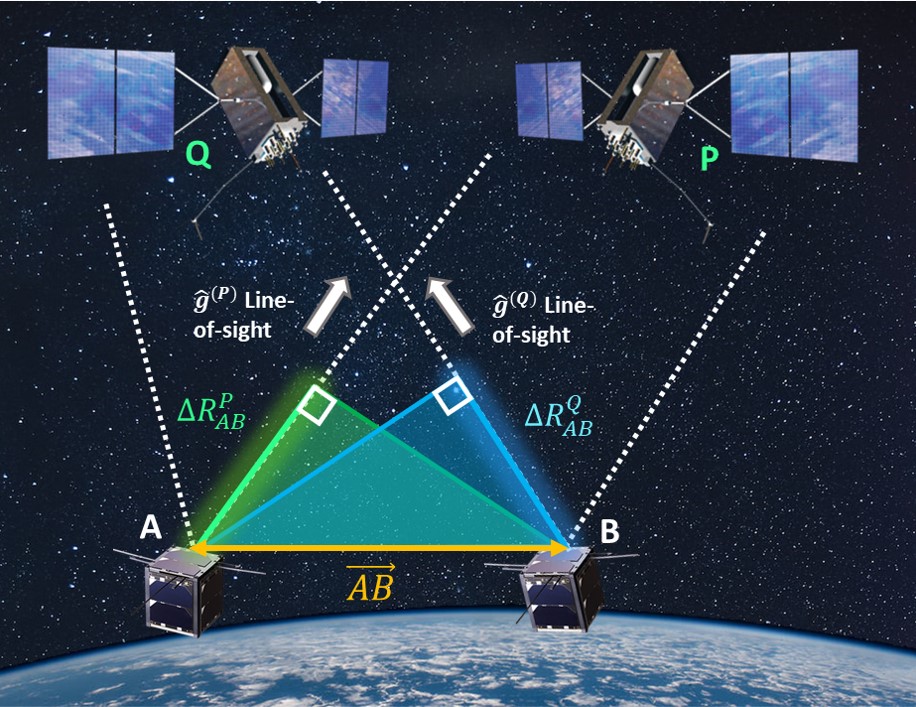}
    \caption{Geometry of double-difference carrier phase}
    \label{figure:review-ddcp}
\end{figure}

where the carrier phase between $A$ and $P$ in distance units $\lambda \phi_A^{(P)}$ is the sum of the geometric range $R_A^{(P)}$, offset by ambiguous integer cycles $N_A^{(P)}$, and corrupted by: an ionospheric delay $I$, the receiver clock bias $c \delta t_A$, GPS satellite clock bias $c \delta t^{(P)}$ from $P$, phase wind-up effects $W^{(P)}$, and thermal noise $\varepsilon_A^{(P)}$. For a detailed analysis of carrier phase measurement error budgeting, the reader is invited to peruse \cite{gpstextbook2006} and the work of Psiaki et al \cite{psiaki2007cdgps}. For cooperative spacecraft with an inter-satellite cross link, these measurements can be communicated to each other and differenced. This Single Difference (SD) operation is defined by $\Delta \left( \cdot \right)_{AB}^{(P)} = \left( \cdot \right)_{A}^{(P)} - \left( \cdot \right)_{B}^{(P)}$. As a result, correlated errors over a short baseline $I$, $c \delta t^{(P)}$ and $W^{(P)}$ cancel out in the SD operation, resulting in a Single Difference Carrier Phase (SDCP) measurement as

\begin{equation}
\label{eq2}
  \lambda \Delta \phi_{AB}^{(P)} =
  \Delta R_{AB}^{(P)} + \lambda \Delta N_{AB}^{(P)} + 
  c \Delta \delta t_{AB} + \Delta \varepsilon_{AB}^{(P)}
\end{equation}

Geometrically relating equation \ref{eq2} with Figure \ref{figure:review-sdcp}, $\Delta R_{AB}^{(P)}$ forms the base of a right-angled triangle with the baseline $\overrightarrow{AB}$ as the hypotenuse. One may further take differences of two SDCP measurements between GPS satellites $P$ and $Q$. This is the Double Difference (DD) operator defined by $\nabla \Delta \left( \cdot \right)_{AB}^{(PQ)} = \Delta \left( \cdot \right)_{AB}^{(P)} - \Delta \left( \cdot \right)_{AB}^{(Q)}$. Common differential receiver clock biases are eliminated in the Double Difference Carrier Phase (DDCP) measurement model given by

\begin{equation}
\label{eq3}
  \lambda \nabla \Delta \phi_{AB}^{(PQ)} =
  \nabla \Delta R_{AB}^{(PQ)} +
  \lambda \nabla \Delta N_{AB}^{(PQ)} + 
  \nabla \Delta \varepsilon_{AB}^{(PQ)}
\end{equation}

Geometrically relating the equations again, $\nabla \Delta R_{AB}^{(PQ)}$ is a linear combination of the baseline $\overrightarrow{AB}$ projected onto each line-of-sight unit vector as per Figure \ref{figure:review-ddcp}. The lines-of-sight $\hat{g}^{(P)}$ and $\hat{g}^{(Q)}$ are known from the GPS satellite antenna and an estimated receiver antenna position. The DDCP measurement model can be re-arranged from equation \ref{eq3} into a least-squares solvable form given multiple measurements,

\begin{equation}
\begin{split}
  \lambda \nabla \Delta \phi_{AB}^{(PQ)} & = 
  \left( \hat{g}^{(P)} - \hat{g}^{(Q)} \right)
  \cdot \overrightarrow{AB} \\
  & + \nabla \Delta N_{AB}^{(PQ)} + \nabla \Delta \varepsilon_{AB}^{(PQ)}
\end{split}
\label{eq4}
\end{equation}

where the baseline $\overrightarrow{AB}$ and $\nabla \Delta N_{AB}^{(PQ)}$ DDCP ambiguities are unknown states that can be solved by least squares (with multiple measurements of equation \ref{eq4}) or by sequential filtering. With DDCP measurement noise typically at the mm-level, fixing the float ambiguities into integers with sufficient certainty allows the filter to converge the state estimate of baseline coordinates towards an equivalent precision of the thermal noise. This illustrates the critical reliance of precise relative navigation on successful IAR.

From this point onwards, the notation $\nabla \Delta \left( \cdot \right)_{AB}^{(PQ)}$ is dropped for brevity when describing ambiguities. Thus, the ambiguity vector $N$ refers to DDCP ambiguities. The tilde notation $\Tilde{N}$ is introduced to refer to DDCP float ambiguities. The most well-established approach for IAR is done via minimization of the Integer Least Squares (ILS) objective using the Least Squares Ambiguity Decorrelation Adjustment, or LAMBDA \cite{teunissen1994lambda} method described below. IAR implementation in DiGiTaL \cite{giralo2019digital} \cite{giralo2021digital} \cite{visors2023aas} is detailed in Section \ref{section5}.

A discrete search for the integers must be performed due to the integer nature of the ambiguities \cite{teunissen1994lambda}. As DDCP measurements are highly correlated due to a common single-difference reference measurement from GPS satellite $Q$, LAMBDA reduces the size of this search space by decorrelating the ambiguities using the integer-preserving Z-transform which begins with an LDL decomposition of the DDCP float ambiguity covariance matrix $Q_{\Tilde{N}}$

\begin{equation}
\label{eq5}
  Q_{\Tilde{N}} = LDL^T
\end{equation}

where $L$ is a lower uni-triangular matrix and $D$ is a matrix of positive diagonals, obtained from the decomposition of the symmetric positive definite covariance $Q_{\Tilde{N}}$. The matrix $D$ in particular is useful as a measure of the variance of float ambiguities after decorrelation. The Z-transform matrix $Z$ is an integer approximation of $L$ that fulfils the requirements given in \cite{teunissen1995invertible}. $Z$ is then applied to the distribution of $n$ float DDCP ambiguities $\Tilde{N}$ for decorrelation,

\begin{align}
\label{eq6}
  Q_{\Tilde{N}_z} & = Z^{T} LDL^T Z \approx D \\
  \Tilde{N}_z & = Z^{T} \Tilde{N} \nonumber
\end{align}

The result is a mean and covariance that is nearly decorrelated (not completely due to the integer constraints). Next, $\Tilde{N}_z$ undergoes recursive conditional rounding $N_{Z}^{(B)} \leftarrow \Tilde{N}_z$ using the integer bootstrapping technique \cite{teunissen1998success}, where $[ \cdot ]$ denotes a rounding operator, subscript $(Z,i)$ indicates the $i^{th}$ ambiguity in the vector of Z-transformed ambiguities, and $I = 1, \ 2, \ ..., \ i-1$ are indices of previously rounded ambiguities, in the equations

\begin{equation}
\label{eq16}
\begin{split}
    N_{Z,1}^{(B)} & = \left[ \Tilde{N}_{Z,1} \right] \\
    N_{Z,2}^{(B)} & = \left[ \Tilde{N}_{Z,2|1} \right]
    = \left[ \Tilde{N}_{Z,2}
    - \frac{\sigma_{2,1}}{\sigma_1^2} \left(
    \Tilde{N}_{Z,1} - N_{Z,1}\right) \right] \\
    \vdots \quad & \qquad \qquad \vdots \\
    N_{Z,n}^{(B)} & = \left[ \Tilde{N}_{Z,n|N}] \right]
    = \left[ \Tilde{N}_{Z,n}
    - \sum_{i=1}^{n-1} \frac{\sigma_{n,i|I}}{\sigma_{i|I}^2} \left( \Tilde{N}_{Z,i|I} - N_{Z,i}\right) \right] \\
\end{split}
\end{equation}

while the $\sigma_{i,j}$ terms in the bootstrapping equations \ref{eq16} are elements of the lower-triangular $L$ and diagonal $D$ matrices resulting from LDL decomposition as follows

\begin{equation*}
\label{eq17}
\footnotesize
L = \\
\begin{bmatrix}
    1 & & & \\
    \frac{\sigma_{2,1}}{\sigma_1^2} & 1 & & \\
    \vdots & \ddots & \ddots & \\
    \frac{\sigma_{n,1}}{\sigma_1^2} & \cdots &
    \frac{\sigma_{n,n-1|N-1}}{\sigma_{n-1|N-1}^2} & 1
\end{bmatrix}
, \quad 
D = \\
\begin{bmatrix}
    \sigma^2_{1} & & & \\
    & \sigma^2_{2|1} & & \\
    & & \ddots & \\
    & & & \sigma^2_{n|N} \\
\end{bmatrix}
\end{equation*}

The success rate of the bootstrapped integers was proven to be greater than or equal to that of simple integer rounding \cite{teunissen1998success}. Thus, using the bootstrapped integers as an initial point to begin the integer search is favoured over using rounded floats. The Mahalanobis distance $\chi$ between float-to-bootstrapped integers provides a reasonable search-width

\begin{equation}
\label{eq19}
  \chi = || N_z^{(B)} - 
  \Tilde{N}_z ||^2_{Q^{-1}_{\Tilde{N}_z}}
\end{equation}

An integer search that minimizes the objective function is then performed over the decorrelated space,

\begin{equation}
\label{eq7}
  \min_{N_z} || N_z - 
  \Tilde{N}_z ||^2_{Q^{-1}_{\Tilde{N}_z}}
\end{equation}

where $N_z$ is the candidate (Z-transformed) vector of integers. When the best candidate can no longer be improved, it is only resolved into an integer if it passes the closed-form success rate test in equation given by 

\begin{equation}
\label{eq8}
  \text{P(success)} = 
  \prod_{i=1}^{n}
  \sqrt{1 - \exp{\left( - 
  \frac{1}{8d_i^2} \right)}} > \kappa_P
\end{equation}

where $n$ is the number of tracked DDCP ambiguities, $d_i$ are $i^{th}$ diagonals of the matrix $D$ in equation \ref{eq8}. It must also pass a discrimination test

\begin{equation}
\label{eq9}
  \norm{ N^{\dagger}_z - 
  \Tilde{N}_z }^2_{Q^{-1}_{\Tilde{N}_z}}
  \bigg/
  \norm{ N^{\ddagger}_z - 
  \Tilde{N}_z }^2_{Q^{-1}_{\Tilde{N}_z}}
  > \kappa_D
\end{equation}

where $N^{\dagger}_z$ and $N^{\ddagger}_z$ in equation \ref{eq9} are the best and second best vector of candidates found. The success thresholds $\kappa_P = 99\%$ and $\kappa_D = 3$ are commonly used in literature \cite{teunissen1998success}.

\newpage


\section{Literature Review}
\label{section4}

Literature which specifically addresses real-time IAR for high-precision in-orbit operations under adverse noise is limited, which is unsurprising considering the challenge of real-time in-orbit IAR per se. In fact, to the best of the authors' knowledge, the VISORS mission will be the first to attempt this \cite{visors2023aas}, due for launch in October 2024 on the SpaceX Transponder-12. This gap in literature and practice is an opportunity for this paper's contributions. The nature of the noise sources listed in Section \ref{section2} prompted this study to segregate the literature review into two broad directions. The first track explores existing on-board metrologies for sensor fusion in order to overcome the innate noise `barriers' in a noisy signal environment. The second track reviews how new integer search and optimization techniques have evolved in the state of the art, and investigates if they can be adapted for rapid real-time integer resolution under a rapidly changing GPS/GNSS geometry. 


\subsection{Reviewing Sensor Fusion for In-Orbit IAR}

Sensor fusion is a commonly adopted practice in literature for automotive and robotics applications requiring the circumvention of poor GNSS visibility and noise challenges in urban canyons. A full suite of sensors fused could include range finders, vision-based navigation, and differential GPS with respect to a reference station \cite{won2014gnss}. For in-orbit applications, two independent studies by Renga \cite{renga2013ranging} and Yang \cite{yang2014ranging} have studied the effects of ranging measurements on IAR performance under in-orbit simulations. Both studies employed an EKF with float ambiguities in the state vector, and both had agreeable demonstrations showing that ranging accelerates the filter's estimate of float ambiguities while also smoothing out navigation error anomalies. The coupling of actual laser ranging with CDGPS in-practice is intended to be demonstrated during the VISORS mission \cite{visors2023aas}. Another set of studies investigated the fusion of bearing angles extracted from vision-based navigation and CDGPS with IAR in GEO under high thermal noise, by Capuano et al \cite{capuano2019dgnss} \cite{capuano2022dgnss}. This study demonstrated that using angle measurements from point-registered, LED-aided, features on a close-range target accelerates float ambiguity convergence despite a drastically high thermal noise influence on the received carrier phase of GPS sidelobe signals in GEO. Overall, sensor fusion offers a promising direction towards achieving IAR under harsh operating conditions especially with the inclusion of metrologies agnostic to GPS/GNSS noise conditions.


\subsection{Reviewing Integer Ambiguity Resolution Algorithms}

Integer resolution is often achieved through the minimization of the ILS objective function. LAMBDA has remained state of the art as a means of solving this, since its introduction \cite{teunissen1994lambda} \cite{teunissen1997results}, with little change to the core algorithm. The Integer Least Mean-Squared (ILMS) error was proposed as an alternative objective to ILS, where it was proven that a wider class of integer equivariant estimators were optimal for minimizing ILMS rather than ILS \cite{teunissen2003inteqest}. Still, ILS minimization using LAMBDA-based techniques have been the most common approach in literature for tackling the IAR problem. Variants of LAMBDA, such as the Modified LAMBDA (mLAMBDA) \cite{chang2005mlambda}, have focused on improving computational efficiency by exploiting the structure and symmetry of the matrices. LAMBDA with search constraints garners recent interest to address robustness and accuracy of the integer search. Henkel formulated a variant of LAMBDA with inequality constraints \cite{henkel2011iar}, enforced by augmenting the cost function in \ref{eq7} with a penalty or barrier function. Jurkowski applied this concept of penalties to a baseline defined by a finite tether of known dimensions (length and orientation) for freight stabilization \cite{jurkowski2011iar}, where physical dimensions of the tether formed a-priori constraints on integer search resulting in significant improvements in baseline estimation. To address the problem of limited time visibility of integers and rapid IAR, Teunissen proposes a partial ambiguity resolution (PAR) variant of LAMBDA \cite{teunissen1999par}. Classical LAMBDA resolves integers as a full batch. LAMBDA with PAR relaxes this condition, selecting only a subset of ambiguities that maximizes some metric success. Teunissen proposes the ambiguity dilution of precision (ADOP) as one such metric \cite{teunissen1997adop}. Parkins explores both ADOP and the signal-to-noise ratio as metrics, applying it to single-epoch PAR in terrestrial receivers, thereby circumventing carrier phase cycle slip management \cite{parkins2011par}. Medina fuses both LAMBDA with constraints \cite{henkel2011iar} and LAMBDA with partial resolution \cite{teunissen1999par} by introducing a precision-driven LAMBDA PAR variant where the formal precision of the fixed solution is included in the cost function as a constraint and the partial subset selection is realized based on the projection of the ambiguities into the position domain \cite{medina2021par}. In summary, the application of constraints in the IAR process offers the advantage of either robustifying the integer search using a-priori information, or enforcing the necessary navigation precision requirements in the cost function. The adoption of a partial resolution strategy is shown to accelerate time-to-first-fix integer resolution, which is a significant advantage for a rapidly changing GPS/GNSS geometry, observed particularly in low Earth orbit. These advantages inspired their adoption in this work.


\section{Navigation Architecture}
\label{section5}


\subsection{Overview of Navigation Architecture}

\begin{figure*}[ht!]
	\centering
    \includegraphics[width=0.88\textwidth]{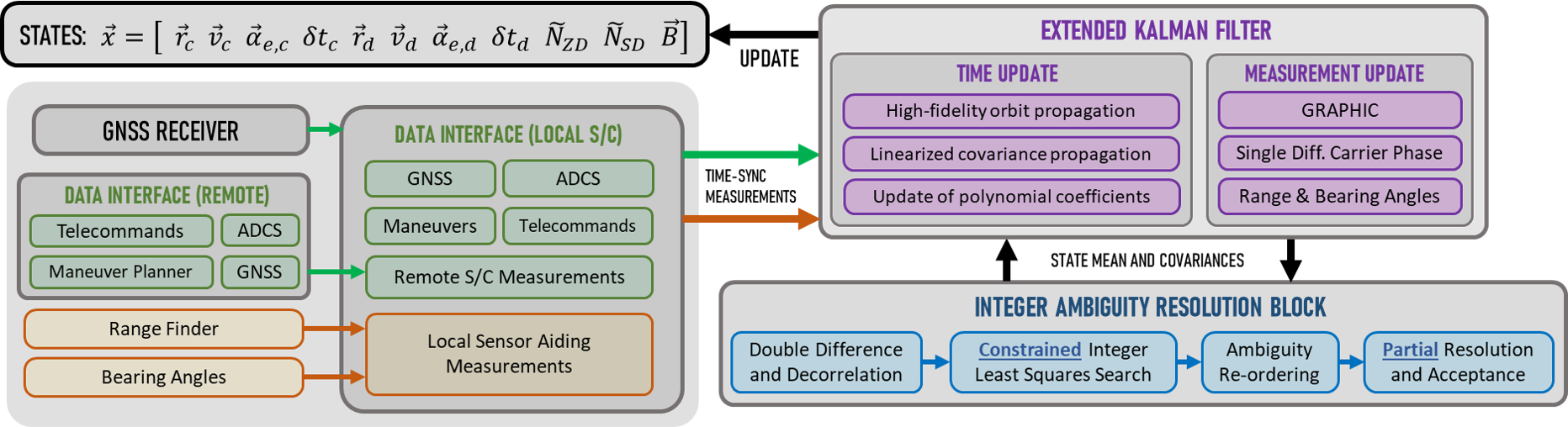}
    \caption{The DiGiTaL navigation flight architecture with three key blocks: (1) the data interface which handles message routing, coordinate transformations, and validation of measurement time-tags of both GPS \textcolor{txtgreen}{(green blocks)} and non-GPS external sensor measurements \textcolor{txtorange}{(orange blocks)}, for both the local spacecraft and a remote partner spacecraft; (2) an orbit determination block driven by an EKF \textcolor{txtpurple}{(purple blocks)}; and (3) the IAR block which fixes, decorrelates, and performs a constrained integer search with partial resolution of DDCP ambiguities \textcolor{txtblue}{(blue blocks)}}
    \label{figure:block-digital}
\end{figure*}

\begin{figure*}[hb!]
	\centering
    \includegraphics[width=0.88\textwidth]{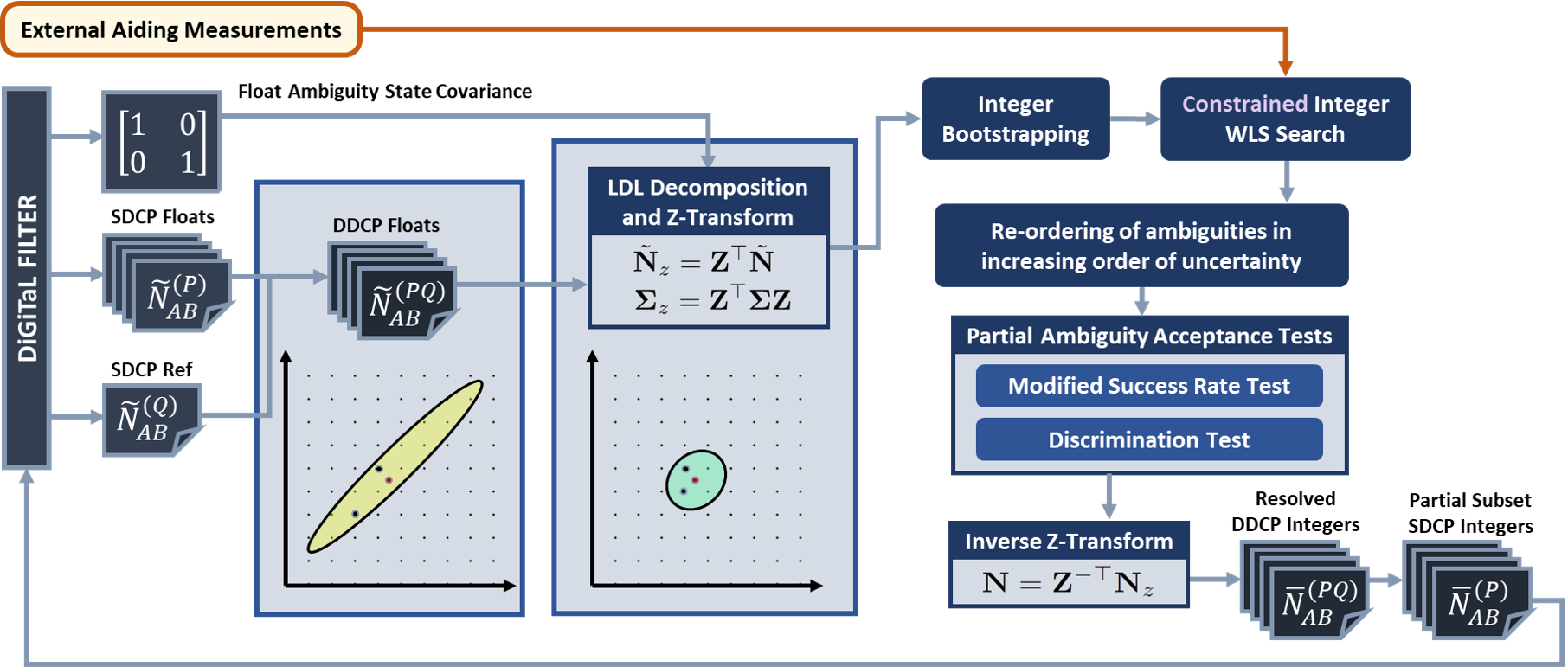}
    \caption{An expanded view of the \textcolor{txtblue}{IAR block} illustrating the sequential steps of integer resolution in DiGiTaL}
    \label{figure:par-lambda}
\end{figure*}

The navigation architecture adopted in this study builds on a tailored variant of SLAB's DiGiTaL flight software \cite{giralo2019digital} \cite{giralo2021digital}. DiGiTaL leverages the powerful error-cancellations of the Group and Phase Ionospheric Calibration (GRAPHIC) measurements \cite{yunck1993graphic} for absolute position estimation, and the Single-Difference Carrier Phase (SDCP) measurements for precise baseline estimation between two cooperative and communicating spacecraft across an inter-satellite link \cite{damico2010thesis}. The filter also accepts other non-GPS external sensor measurements such as range from a rangefinder and bearing-angles from vision-based sensors such as star trackers, at far-range. It is assumed that bearing angles are resolvable through an image processing module \cite{koenig2021artms}.

The navigation filter of each spacecraft tracks both local (chief) and remote (deputy) spacecraft states

\begin{equation}
\label{eq11}
\footnotesize
    \Vec{x} = \left[ 
    \Vec{r}_c, \ \Vec{v}_c, \ 
    \Vec{\alpha}_{e,c}, \ c \delta t_c,
    \Vec{r}_d, \ \Vec{v}_d, \ 
    \Vec{\alpha}_{e,d}, \ c \delta t_d,
    \Tilde{N}_{ZD}, \Tilde{N}_{SD}, \ \Vec{B} \right]
\end{equation}

where $\Vec{r}_c$, $\Vec{v}_c$, $\Vec{r}_d$, $\Vec{v}_d$, are the Earth-centered inertial (ECI) positions and velocities of the chief and deputy spacecraft center of mass respectively. Dynamic model compensation through the estimation and application of stochastic empirical acceleration terms $\Vec{\alpha}_{e,c}$, $\Vec{\alpha}_{e,d}$ account for unmodelled dynamics \cite{stacey2021adaptive}. These are applied as perturbative forces in the radial-tangential-normal (RTN) frame of the chief (or target), and added to the force model of the orbit propagation step in the filter time update, which is a numerical orbit propagation using the GRACE GGM-05S model of degree and order 20, plus the empirical accelerations. For GPS L1, the receiver clock offset of the chief $c \delta t_c$ and deputy $c \delta t_d$ are scalars. Receiver clock offsets and empirical accelerations are modelled as first-order Gauss-Markov processes. Receiver clock offsets are prescribed an infinite time constant in order to describe a random walk process \cite{damico2010thesis}. States $\Tilde{N}_{ZD}$ and $\Tilde{N}_{SD} \in \mathbb{R}$$^{24}$ are the undifferenced and SDCP float ambiguities. $\Tilde{N}_{SD}$ floats are resolved into integers only upon satisfying acceptance tests \cite{teunissen1998success}, as depicted in Figure \ref{figure:par-lambda}. The filter also tracks (possible) sensor biases of the external sensor measurements. These are captured in a $3 \times 1$ state $\Vec{B}$ for range, azimuth and elevation angle biases as seen in the local sensor frame (see Figure \ref{figure:angles-ref-frame} in next section). The total state vector size is $\Vec{x} \in \mathbb{R}$$^{71}$.

The time update of the navigation filter is not a direct update to the filter states, but rather an update to the coefficients of a 5th order Hermite polynomial fitted over the states to allow for on-demand interpolation at a much higher time resolution. This approach was successfully flown on the BIRD \cite{briess2005bird} and PRISMA \cite{damico2010thesis} missions. All filter parameters are listed in the Appendix, Table \ref{tab:filter-params}.

The principal contribution to state-of-the-art in this paper is an integrated three-step architecture for achieving IAR under adverse signal-to-noise conditions, through sensor and data fusion: (i) a \textit{loose coupling} stage where non-GPS external sensor measurements are fused with CDGPS measurements in the Kalman filter measurement update; (ii) a \textit{tight coupling} stage where the external sensor measurements are directly incorporated into the ILS minimization in order to assess the quantified agreement between the integer candidates and the external sensor measurements; and (iii) a partial resolution step with modified acceptance tests that empirically accounts for the influence of this quantified agreement on the success rate. This agreement is evaluated for each integer candidate during the search. The result is a comprehensive navigation architecture that integrates CDGPS measurements with sensor coupling through a multi-stage process, packaged into a unified and flight-capable software in C++. Such an architecture has yet to be proposed in literature, much less validated through high-fidelity simulations. This is the key contribution behind this work.

Additionally, another critical innovation that enables the practicality of this architecture is its ability to balance computational efficiency with navigation precision. With a covariance matrix of size $71 \times 71$, matrix operations necessitate computational optimizations such as sparse matrix operations, leveraging symmetry during matrix decompositions, and online resizing of the covariance matrix through dynamic programming whenever possible. These optimizations enable rapid online state updates with IAR on a flight computer at an estimated cadence $< 30$ seconds \cite{visors2023aas}.


\subsection{Loose-Coupling Implementation}

The motivation for the loose coupling stage arises from the prevalence of on-board sources of metrologies which are agnostic to GPS/GNSS signal noise conditions \cite{renga2013ranging} \cite{yang2014ranging} \cite{capuano2019dgnss} \cite{capuano2022dgnss}. On-board cameras and star-trackers provide angles-only measurements of targets typically at $\approx 100$ arcsec precision, after measurement assignment in resolved images. The ARTMS software on-board the NASA Starling mission is a prime example \cite{koenig2021artms}. Range measurements can be embedded in the cross-link via asymmetric two-way ranging \cite{jiang2007ranging}, or through a laser-range finder which offers mm-level precision \cite{visors2023aas}. A joint filter measurement update with CDGPS measurements are then performed. 

It is important to note that any improvement to the filter's distribution about the float ambiguities $\Tilde{N}_{ZD}$ and $\Tilde{N}_{SD}$ happens only indirectly from the external measurements, since ambiguities are not observable from these sensor measurements. The distribution of ambiguities $\Tilde{N}_{ZD}$ and $\Tilde{N}_{SD}$ improve because there is an improvement in the estimation of baseline coordinates $\Vec{r}_c$ and $\Vec{r}_d$, which are correlated with ambiguities; hence the term \textit{`loose'} coupling.

The first practical consideration in implementation is that the introduction of new sensors may introduce biases that may degrade or evolve over time once in-orbit if not calibrated during flight. This necessitates the estimation of sensor biases $\Vec{B} = [B_R, B_\alpha, B_\varepsilon]$ in the state, as seen in equation \ref{eq11}.

\begin{figure}[ht]
	\centering
    \includegraphics[width=0.5\textwidth]{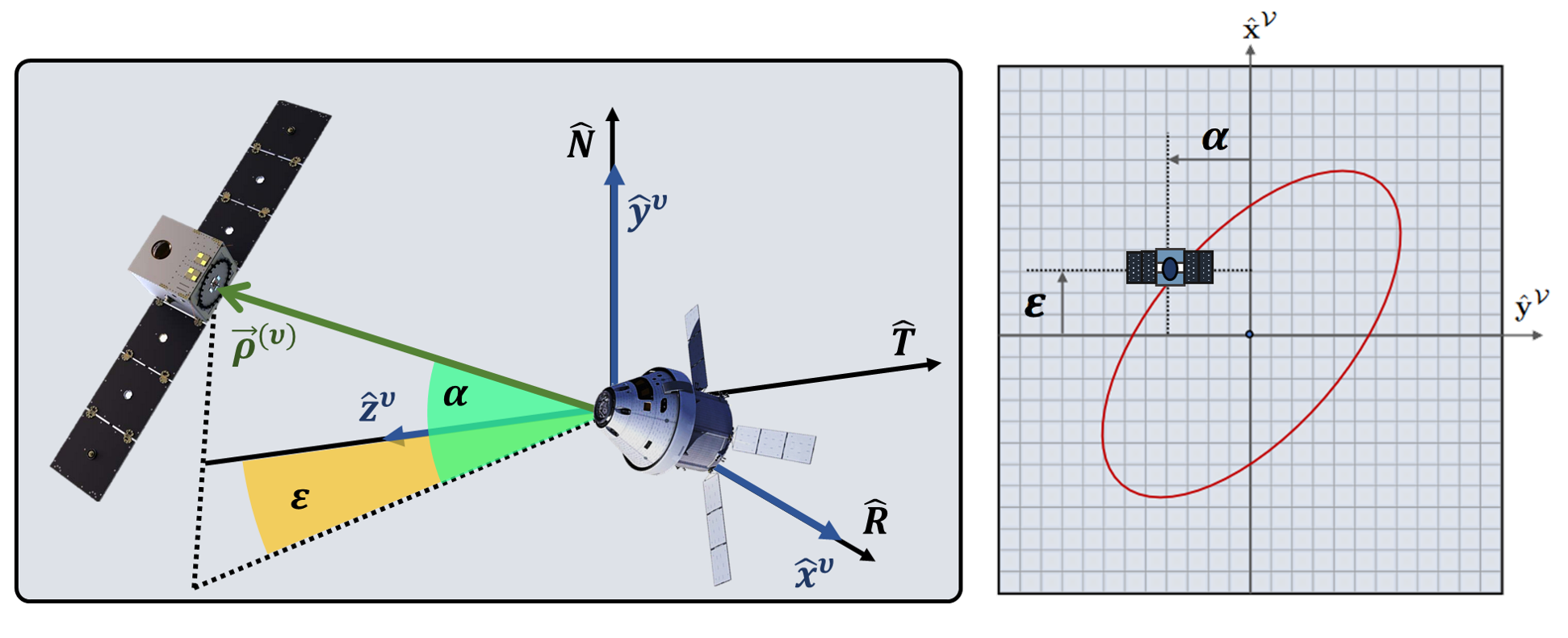}
    \caption{Bearing angles subtend the line-of-sight vector in the vision-based sensor frame, $\upsilon$ (left), and localize the target pixel cluster in the image plane (right)}
    \label{figure:angles-ref-frame}
\end{figure}

The second consideration is that incorporating sensor measurements observed in different frames requires accurate, time-tagged, attitude information in order to transform coordinates between each sensor reference frame. For example, GPS measurements are typically applied in an Earth-centered Earth-fixed (ECEF), whereas vision-based or ranging measurements are taken in a local boresight-aligned sensor frame, as depicted in Figure \ref{figure:angles-ref-frame}. Thus, it is crucial that the attitude time-tag matches the time-tags of measurements, possibly through interpolation, so that measurements are commensurable across different coordinate systems when applying the filter measurement update.

The GRAPHIC measurement is an ionospheric-free linear combination of undifferenced code pseudorange $\rho_{range}$ and carrier phase $\lambda \phi_A^{(P)}$ between the antenna phase centers of receiver A and transmitting GPS satellite P, which eliminates a significant fraction of the code delay and phase advance of the GPS L1 signals through the ionospheric medium,

\begin{equation}
  \rho_{graphic} =
  \frac{1}{2} \rho_{range} +
  \frac{1}{2} \lambda \phi_A^{(P)}
\end{equation}

while the SDCP measurement model from equation \ref{eq2} is re-expressed here for convenience

\begin{equation}
  \lambda \Delta \phi_{AB}^{(P)} =
  \Delta R_{AB}^{(P)} + \lambda \Delta N_{AB}^{(P)} + 
  c \Delta \delta t_{AB} +
  \Delta \varepsilon_{AB}^{(P)}
  \tag{2}
\end{equation}

For modelling the external sensor measurements in this study, it is assumed that both the rangefinder and camera share the same boresight unit vector and thus the same local sensor coordinate frame $\upsilon$ as depicted in Figure \ref{figure:angles-ref-frame}, with coordinate axes [$\hat{x}^{\upsilon}, \ \hat{y}^{\upsilon}, \ \hat{z}^{\upsilon}$]. It is also assumed that there are no sensor mounting boresight errors, and the body-frame mounting offset from the center of mass is known. Let $\Vec{\rho}^{(\upsilon)}$ be the relative position vector from chief to deputy in the $\upsilon$ frame, and let $|| \cdot ||$ denote the L2 norm operator. Then, the range and angles-only measurement models are

\begin{equation}
\label{eq12}
    \bar{R} = \norm{\Vec{\rho}^{(\upsilon)}} + B_R
\end{equation}

\begin{equation}
\label{eq13}
    \bar{\alpha} = \text{sin}^{-1} 
    \left( \frac{\Vec{\rho}^{(\upsilon)} 
    \cdot \hat{y}^{\upsilon}}
    {\norm{\Vec{\rho}^{(\upsilon)}}}
    \right) + B_{\alpha}
\end{equation}

\begin{equation}
\label{eq14}
    \bar{\varepsilon} = \text{tan}^{-1} 
    \left( \frac{\Vec{\rho}^{(\upsilon)}
    \cdot \hat{x}^{\upsilon}}
    {\Vec{\rho}^{(\upsilon)}
    \cdot \hat{z}^{\upsilon}}
    \right) + B_{\varepsilon}
\end{equation}

where the bar notation ($\bar{\cdot}$) denotes computed measurements as opposed to observed measurements (no bar). As an implementation detail, it must be noted that the Jacobians of the observations $\mathbf{H}_R$, $\mathbf{H}_\alpha$ and $\mathbf{H}_\varepsilon$ are taken with respect to components of [$\hat{x}^{\upsilon}, \ \hat{y}^{\upsilon}, \ \hat{z}^{\upsilon}$] in the sensor frame, and thus care must be taken when applying the Kalman filter measurement updates to states in equation \ref{eq11} that are expressed in the ECI frame. The observation Jacobian $\mathbf{H}_i$ (where $i \in [R, \alpha, \varepsilon]$) must be transformed by the $3 \times 3$ direction cosine matrices $\mathbf{\Theta}_{ECI}^{\upsilon}$ mapping ECI frame coordinates to sensor frame coordinates,

\begin{equation}
\label{eq15}
    \mathbf{H}_i^{ECI} = \mathbf{H}_i \
    \mathbf{\Theta}_{ECI}^{\upsilon}
\end{equation}

which again underlines the importance of accurately time-tagged attitude coordinates in order to prevent coordinate transformation errors in both the measurements and their Jacobians $\mathbf{\Theta}_{ECI}^{\upsilon}$. The sensor Jacobians in the $\upsilon$-frame are appended in the Appendix.

For simulation purposes, the sensor noise budgets in Table \ref{tab:ext-error-budget} below are assumed throughout the study,

\begin{table}[h]
\renewcommand{\arraystretch}{1.3}
\caption{\bf External aiding sensor error budget}
\label{tab:ext-error-budget}
\centering
\begin{tabular}{|c|c|c|c|}
    \hline
    Measurements & $R$ & $\alpha$ & $\varepsilon$ \\
    \hline \hline
    Bias & 50 mm & 500 arcsec & 500 arcsec \\
    \hline
    Std. Dev. & 5 mm & 100 arcsec & 100 arcsec \\
    \hline
\end{tabular}
\end{table}

A back-of-the-envelope calculation can be done to assess the impact of external sensor measurements on IAR success rate, assuming that the bias state vector $\Vec{B}$ converges to the correct value. Consider the steady-state covariance $\mathbf{\Sigma_\infty}$ of tracked SDCP ambiguities $\Tilde{N}_{SD}$, assumed to be uncorrelated and thus having a diagonal covariance. Then, $\mathbf{\Sigma_\infty}$ at steady-state can be approximated by solving the quadratic Discrete-Time Algebraic Riccatti Equation (DARE) for the linear Kalman filter (only for approximation purposes),

\begin{equation}
\begin{split}
    \mathbf{\Sigma_\infty} & = \mathbf{A}
    \Sigma_\infty \mathbf{A^T} + \mathbf{Q} \\
    & - \mathbf{A} \Sigma_\infty \mathbf{C^T}
    \left( \mathbf{C} \Sigma_\infty
    \mathbf{C^T} + \mathbf{R} \right)^{-1}
    \mathbf{C} \Sigma_\infty \mathbf{A^T}
\end{split}
\end{equation}

where linear time-invariance is assumed for only $\Tilde{N}_{SD}$ states, the state transition matrix $\mathbf{A}$ is the identity matrix of the appropriate dimensions, with a sensitivity matrix $\mathbf{C} = \lambda \mathbf{1}^T$, as per the SDCP measurement model equation \ref{eq2}, and $\mathbf{Q} = I \times$$10^{-3}$ cycles. Diagonals of $\mathbf{\Sigma_\infty}$ are applied into the success rate test given in equation \ref{eq8}, assuming 10 float ambiguities, and GPS L1 $\lambda = 19.05$cm to produce the following back-of-the-envelope estimates for the IAR success rate,

\begin{table}[h]
\renewcommand{\arraystretch}{1.3}
\caption{\bf Back-of-the-envelope IAR success rates at steady-state, for 10 tracked SDCP float ambiguities}
\label{tab:back-of-the-envelope}
\centering
\begin{tabular}{|c|c|c|c|}
    \hline
    \bfseries R & $\lambda/4$ & $\lambda/8$ & $\lambda/40$ \\
    \hline \hline
    Success Rate & $28.17 \%$ & $51.79 \%$ & $95.41 \%$ \\
    \hline
\end{tabular}
\end{table}

The first scenario with $\mathbf{R}$ $=\lambda / 4$ represents the CDGPS measurements under the theoretical maximum carrier phase noise due to multi-path influence \cite{svehla2018leo}, and no external sensor measurements are applied; the second scenario with $\lambda / 8$ represents a user-equivalent range-error using bearing angles, with errors scaled to the arc-length at a 50m range, following the error budget in Table \ref{tab:ext-error-budget}; the third scenario with $\lambda / 40$ represents a user-equivalent range-error when applying a precise laser ranging measurement following Table \ref{tab:ext-error-budget}. Flight-like simulation results detailed later in Section \ref{section7} would support the idea that external measurements do indirectly increase the success rate of IAR in the loose coupling step.


\subsection{Tight-Coupling Implementation Details}

The motivation for the \textit{tight coupling} stage is to provide an additional data editing step during integer search and resolution by directly incorporating the external aiding measurements in the choice of candidate integer ambiguities. The initialization of tight-coupling is similar to LAMBDA where the Z-transform is applied on the distribution (mean and covariance) of the DDCP ambiguities as per equations \ref{eq5} and \ref{eq6}, followed by integer bootstrapping in order to obtain an initial guess and also set the search width of the integer space.

From here, the tight-coupling algorithm begins the search for the optimal DDCP ambiguity vector using a local best-neighbour search, also known as the Hooke-Jeeves method (see Algorithm 7.5 in \cite{kochenderfer2019algorithms}), with a pre-defined step size $k$ per iteration. Each step taken is in terms of the number of ambiguity cycles from the previous vector of integer candidates. The search evaluates the total cost at the current choice of integer candidates, and then evaluates the cost for each step in $\pm k$ directions, accepting the best improvement it finds for each ambiguity. If no improvements are found, the step size decreases by 1 cycle, and the process continues until $k=0$. The cost function evaluated at each iteration is the sum of two individual costs. The first cost term is the weighted least squares cost $C_N$ of integer selection, identical to the cost function \ref{eq7} in the original LAMBDA

\begin{equation}
\label{eq20}
    C_N = \norm{N_z - \Tilde{N}_z}^2_{Q^{-1}_{\Tilde{N}_z}}
\end{equation}

while the second cost term is the penalty cost quantifying how much the selected candidate set of integer ambiguities violates the suggested baseline based on the external aiding sensor measurements,

\begin{equation}
\label{eq22}
    C_{ext} = 
    \frac{(\alpha - \bar{\alpha})^2}
    {R \sigma_\alpha^2} +
    \frac{(\varepsilon - \bar{\varepsilon})^2}
    {R \sigma_\varepsilon^2} +
    \frac{(R - \bar{R})^2}
    {\sigma_R^2}
\end{equation}

where $[R, \ \alpha, \ \varepsilon]$ are the observed aiding measurements; $[\bar{R}, \ \bar{\alpha}, \ \bar{\varepsilon}]$ are the computed aiding measurements; $\sigma_\alpha$, $\sigma_\varepsilon$, and $\sigma_R$, are the sensor noise sigmas of the bearing angle and ranging sensors which are assumed known due to pre-flight calibration. The costs of violating angular constraints are weighted by the inter-satellite range $R$ due to the arc-length error that scales linearly with range. 

Since there is no explicit means of obtaining the computed external aiding measurements from the DDCP equations as a function of each candidate DDCP ambiguity vector, $[\bar{R}, \ \bar{\alpha}, \ \bar{\varepsilon}]$ quantities are derived from obtaining the candidate baseline solution from each candidate set of DDCP ambiguities during integer search. The baseline can be solved in closed-form via the least-squares solution of a system of DDCP observations, where each observation is given by equation \ref{eq4}. In matrix form, the DDCP geometry matrix is $\mathbf{G}$, while $\mathbf{\Phi}$ refers to the vector of DDCP measurements, and $\mathbf{N_{cand}}$ refers to the candidate set of ambiguities in the current search iteration

\begin{equation}
\label{eq21}
    \bar{\rho} = (\mathbf{G}^T \mathbf{G})^{-1} 
    \mathbf{G}^T \lambda (\mathbf{\Phi} - N_{cand})
\end{equation}

The computed candidate baseline $\bar{\rho}$ in equation \ref{eq21} is expressed in ECI and must be transformed into the basis of the external aiding sensor frame $\bar{\rho}^{(\upsilon)} \leftarrow \bar{\rho}$ using the direction cosine matrix $\mathbf{\Theta}_{ECI}^{\upsilon}$. The computed range and/or bearing angles $[\bar{R}, \ \bar{\alpha}, \ \bar{\varepsilon}]$ are obtained using the measurement models in equation \ref{eq12}, \ref{eq13}, \ref{eq14}, and compared with the observed range and/or bearing angles $[R, \ \alpha, \ \varepsilon]$ in the cost function \ref{eq22} as the norm-squared, noise-weighted sum of observed-minus-computed residuals. The total cost evaluated during the integer search is simply the sum of equations \ref{eq20} and \ref{eq22}, and the minimization is described by

\begin{equation}
\label{eq23}
  \min_{N_z} (C_N + C_{ext})
\end{equation}

\begin{figure}[ht]
    \centering
    \includegraphics[width=0.48\textwidth]{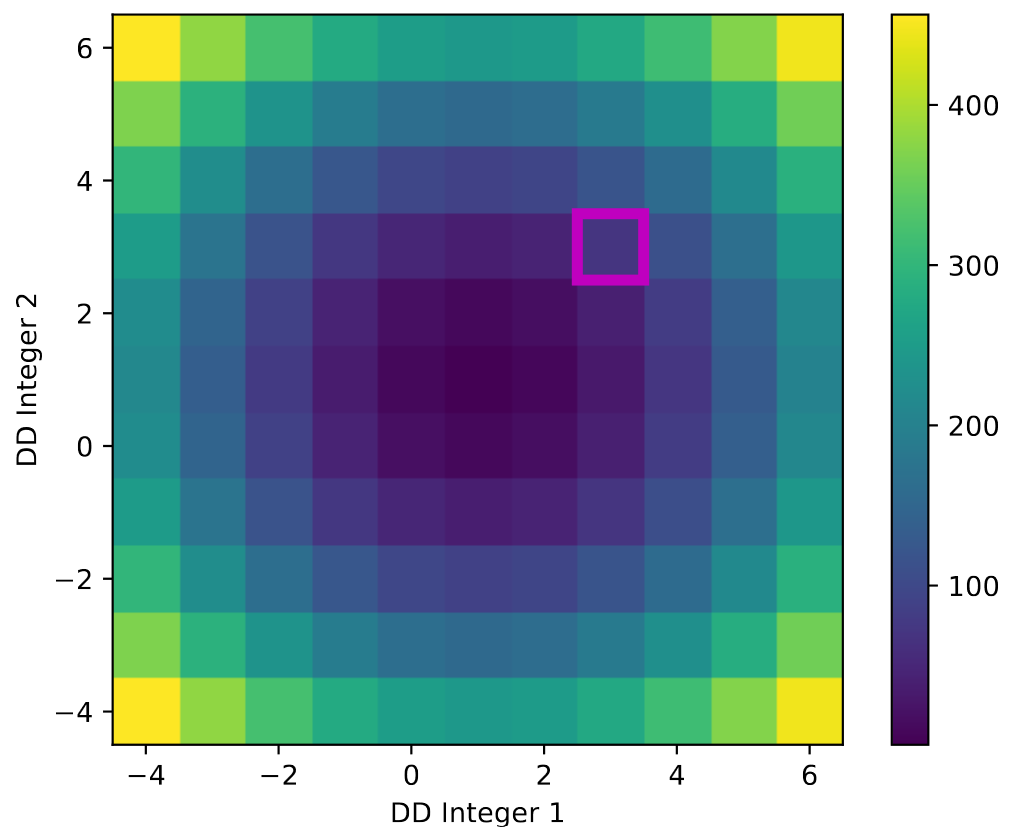}
    \caption{Sampled cost function \textbf{without} constraints (only weighted integer selection costs \ref{eq20} are considered)}
    \label{figure:iar-cost-no}
\end{figure}

\begin{figure}[ht]
    \centering
    \includegraphics[width=0.48\textwidth]{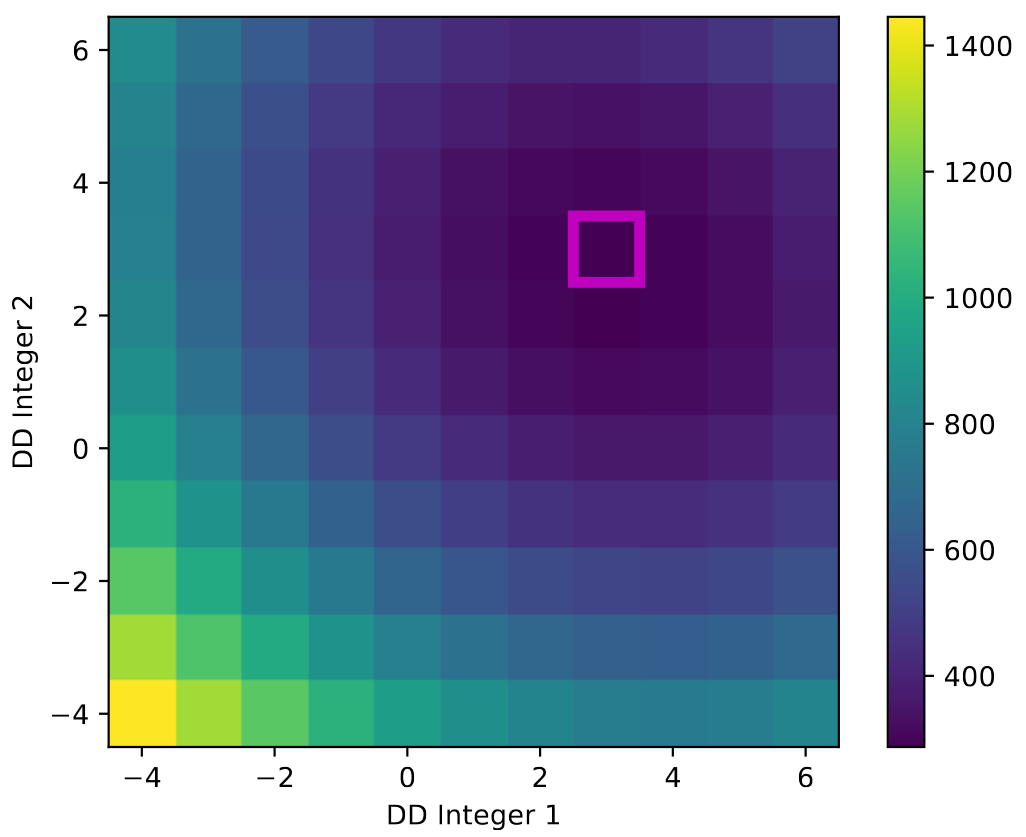}
    \caption{Sampled cost function \textbf{with} constraints (the total constrained cost \ref{eq23} is considered)}
    \label{figure:iar-cost-yes}
\end{figure}

The addition of the penalty cost function in equation \ref{eq22} can be better appreciated with a snapshot of the cost functions with and without constraints. These snapshots are taken from an example simulation under maximum multi-path influence, across an integer subspace spanned by two unresolved DDCP float ambiguities in Figures \ref{figure:iar-cost-no} and \ref{figure:iar-cost-yes},

\newpage

The sample snapshots were taken at an instance where \textit{all} float ambiguities remain unresolved and the theoretical success rate test (in that snapshot) evaluated to $9.37\%$ using equation \ref{eq8} for all $n$ DDCP float ambiguities. Inter-satellite range in this sample was 40.0m. The \textcolor{txtpurple}{$\square$} box is the ground truth DDCP integer tuple. The global minimum of the cost function after a-priori constraints are imposed is shifted much closer to the ground truth value of the DDCP integer ambiguity. This single-epoch example illustrates the potential improvement in accurate integer fixing with the addition of a-priori constraints despite a low P(success) value.

The tight-coupling implementation is outlined in Algorithm \ref{algo:tightcoupling}, which should be executed only after the LDL decorrelation and bootstrapping. The algorithm is initialized with $n$ bootstrapped DDCP ambiguities, $N_z^{(B)}$, and an initial user-defined step size $k$.

\newpage


\subsection{Partial Ambiguity Resolution with Modified Success Rate Test}

The metric of success employed by the partial ambiguity resolution algorithm has been modified from its original form in equation \ref{eq8} to a form that captures the improved robustness of imposing a-priori constraints. The closed-form success rate conditioned on external measurements is not easily found, since as aforementioned, there is no explicit dependency between the distribution of the external sensor measurements and distribution of the integer ambiguities. Thus, an empirical modification to the success rate test is proposed with an added coefficient $S^{\gamma}$ as

\begin{equation}
\label{eq24}
  \text{P(success)} = 
  \prod_{i=1}^{m}
  \sqrt{1 - \exp{\left( - 
  S^{\gamma}
  \frac{1}{8d_i^2} \right)}} > \kappa_P
\end{equation}

\begin{equation}
\label{eq25}
  S^\gamma = \left(
  \frac{C_{best}}
  {C_{init}} \right)^\gamma
\end{equation}

where $m$ is the number of ambiguities in the partial subset, rather than all $n$ unresolved ambiguities; $C_{init}$ and $C_{best}$ are the constrained costs evaluated initially and finally after execution of Algorithm \ref{algo:tightcoupling}. To decide on the number $m$ of candidates selected in the partial subset, the elements in the vector of ambiguities $N_z$ are re-arranged in ascending order of their decorrelated variances after executing Algorithm \ref{algo:tightcoupling}. Then, each ambiguity is successively considered for integer resolution until the entire subset grows in size to the largest possible $m$ value such that it fails to meet the modified success rate test and discrimination test. The subset of fixed integers that pass the acceptance tests will directly replace their previous float estimates with integer values while also zeroing out the state covariances of fixed integers.

\begin{figure}[ht]
	\centering
    \includegraphics[width=0.5\textwidth]{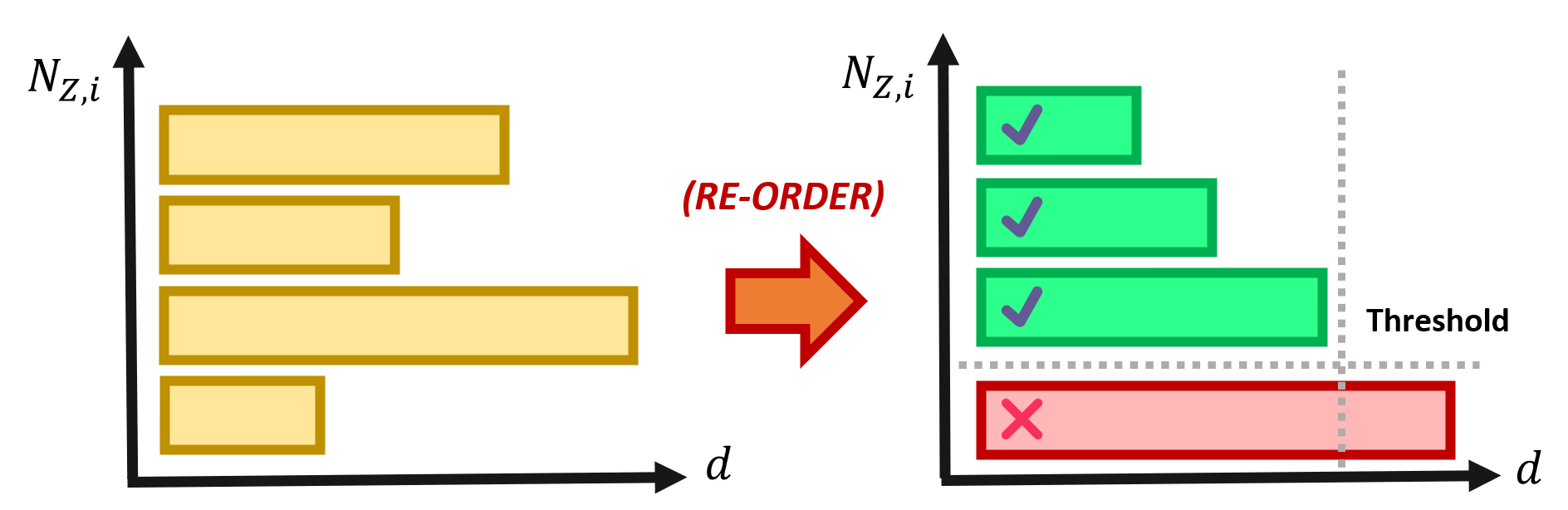}
    \caption{Notional diagram illustrating the re-ordering of DDCP ambiguities in ascending value of their decorrelated covariance diagonals $d_i$ with successive resolution}
    \label{figure:par-ordering}
\end{figure}

The cost improvement coefficient $S$ can be thought of as an empirical measure of the cost savings after executing Algorithm \ref{algo:tightcoupling}, since the diagonals of the decorrelated float ambiguities do not directly reflect the influence of the external aiding sensor measurements in the unmodified success rate test \ref{eq8}. The exponent term $\gamma$ is a float hyper parameter that influences the weight of the coefficient $S$. The naturally intuitive choice of $\gamma = 1/m$ performs well in flight-like simulations, as will be demonstrated in Section \ref{section7}. The threshold success rate $\kappa_P$ remains unchanged at $99\%$ as recommended in literature.

\newpage


\begin{algorithm*}[ht]
\caption{Integer Search under Tight-Coupling}
\label{algo:tightcoupling}
\begin{algorithmic}[1]
\setlength{\parskip}{0.5em}
    \Require $N_z^{(B)}$ and $k$ and external aiding measurements $\left[ R, \alpha, \varepsilon \right]$
    \State $C_{init} \gets C_N + C_{ext}$ \Comment{Initialize cost using equations \ref{eq20} and \ref{eq22}}
    \State $\left[ N_z^{(1)}, N_z^{(2)} \right] \gets N_z^{(B)}$ \Comment{Initialize the 1st and 2nd best candidates}
    \State $\left[ C^{(1)}, C^{(2)} \right] \gets C_{init}$ \Comment{Initialize the 1st and 2nd best candidate costs}
    \While{$k > 0$}
        \State improvements $\gets false$
        \For{$i \in [1, \ 2, \ ... \ n]$} \Comment{For each float ambiguity}
            \For{step $\in [-k,+k]$} \Comment{Step forward and back in the $i^{th}$ dimension}
                \State $N_z^{\text{step}} \gets N_z^{(1)}$
                \State $N_z^{\text{step}}[i] \gets N_z^{\text{step}}[i] +$ step
                \State $N^{\text{step}} \gets$ Inverse Z-transform of $N_z^{\text{step}}$
                \State $\bar{\rho}^{\text{step}} \gets (\mathbf{G}^T \mathbf{G})^{-1} \mathbf{G}^T \lambda (\mathbf{\Phi}$ $- N^{\text{step}})$ \Comment{Equation \ref{eq21}} 
                \State $[\bar{R}, \ \bar{\alpha}, \ \bar{\varepsilon}] \gets$ Get computed (C) external measurements using $\bar{\rho}^{\text{step}}$ \Comment{Equations \ref{eq12} \ref{eq13} \ref{eq14}}
                \State $[R, \ \alpha, \ \varepsilon] \gets$ Get observed (O) external measurements \Comment{Query from sensor}
                \State $C^{\text{step}} \gets C_N + C_{ext}$ \Comment{Equations \ref{eq20}, and \ref{eq22} using O-C residuals}
            \EndFor
            \State Save the best $C^{\text{step}}$ for step $\in [-k,+k]$
            \If{$C^{\text{step}} < C^{(1)}$}
                \State $N_z^{(2)} \gets N_z^{(1)}$
                \State $N_z^{(1)} \gets N_z^{\text{step}}$
                \State $C^{(2)} \gets C^{(1)}$
                \State $C^{(1)} \gets C^{\text{step}}$
                \State improvements $\gets true$
            \EndIf
        \EndFor
        \If{improvements $== false$} 
            \State $k \gets k - 1$ \Comment{Reduce step size if the search did not yield improvements}
        \EndIf
    \EndWhile \\
\Return $\left[ N_z^{(1)}, N_z^{(2)} \right]$
\end{algorithmic}
\end{algorithm*}

\clearpage


\section{Flight Simulation Setup}
\label{section6}

In this section, we validate the proposed technique of loose and tight coupling in order to assess the navigation and IAR performance. Two scenarios are considered here:

\begin{tcolorbox}
\begin{itemize}
  \item \textbf{Scenario 1:} Rendezvous and docking with the International Space Station (ISS) in LEO
  \vspace{2mm}
  \item \textbf{Scenario 2:} Rendezvous and docking with a geostationary micro-satellite in GEO
\end{itemize}
\end{tcolorbox}

The ground truth trajectory is generated in C++ using SLAB's high-fidelity $\mathcal{S}$$^3$ astrodynamics library as per \ref{figure:flight-setup} below. The gravity model uses the GRACE GGM-05S gravity model with a degree and order of 60, in addition to third-body influences by the Sun and Moon; the atmospheric density was modelled using Harris-Priester; and the solar radiation pressure computed is based on the analytical ephemeris of the Sun with a cylindrical shadow model. Trajectories generated in $\mathcal{S}$$^3$ were also exported and visualized in STK (a commercial astrodynamics software),

\begin{figure}[hb!]
	\centering
    \includegraphics[width=0.5\textwidth]{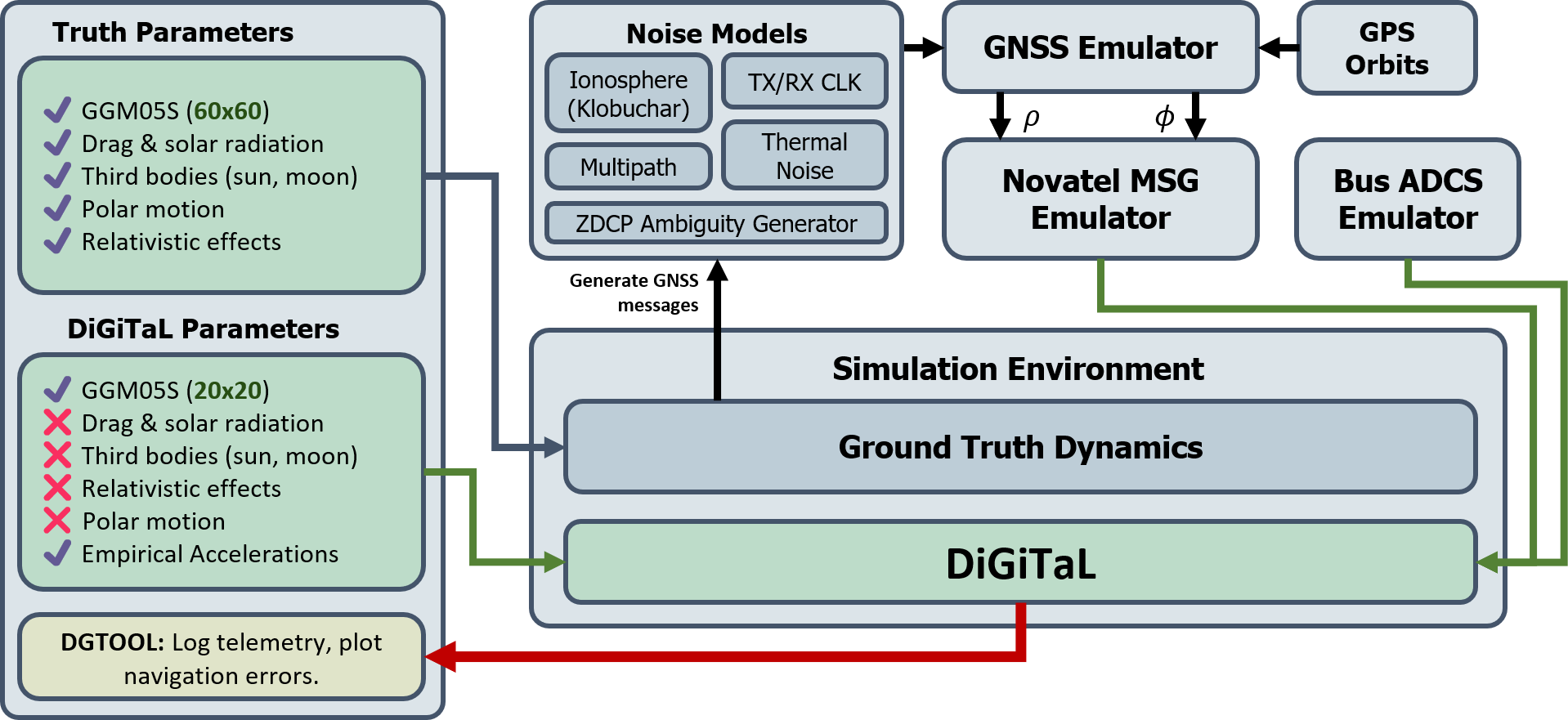}
    \caption{Flight-like simulation setup with DiGiTaL}
    \label{figure:flight-setup}
\end{figure}

For modelling code pseudorange and carrier phase errors, the ionospheric error is modelled via the Klobuchar model with code phase delay equal and opposite to carrier phase advance. Clock errors are modelled as a random walk in distance units, where the random walk's step size per second is sampled from a zero-mean Gaussian with $\sigma_{c \delta t} = 1.0$m. The thermal noise of both code and carrier phase are significant influences on CDGPS and IAR performance as these are non-systematic errors that cannot be cancelled via the linear measurement combinations of GRAPHIC and SDCP \cite{svehla2018leo}. As such, the resultant thermal noise was derived with particular care from the expected $C/N_0$ using a detailed link budget analysis in Table \ref{tab:link-budget}. Factors considered include the receiver internal losses, antenna gains, and the phase-locked-loop (PLL) bandwidth's effects on carrier tracking performance. Specifications from a Novatel OEM628 receiver card were taken as reference, which are applied to a validated error model provided by \cite{psiaki2007cdgps} and \cite{giralo2021digital}.

GPS ephemeris errors are injected as a corruption of orbital elements in the GPS broadcast message using relative orbital elements (ROE). This introduces periodic variations of the Cartesian error as predicted by the mapping between the ROEs and Cartesian coordinates \cite{damicos2006proximity}. ROEs were chosen to enact an equivalent root-mean-square error of $1.5m$ in the GPS ephemeris with zero along-track drift. This method of injecting errors is a more realistic representation of simulating how a broadcast ephemeris' error propagates systematically rather than randomly, until the receiver's next broadcast update.

For modelling multi-path, a simplified Gaussian shadowing model is proposed. The model is by no means a geometrically accurate reflection of the both the local and remote spacecraft's structural influences. It introduces only order-of-magnitude equivalent effects on GPS measurement errors. Two sources of multi-path are distinguished: the near-field multi-path is elevation-dependent and imposed by reflections and shadowing effects cast by the spacecraft's local appendages, as was observed in CHAMP and GOCE missions \cite{svehla2018leo}; the far-field multi-path is imposed by reflections from an external object, such as a partner spacecraft during rendezvous at close range. These two distinct phenomena were demonstrated to be distinguishable through carrier phase measurement residuals before and after the separation of both the Mango and Tango formation flying spacecraft during the PRISMA SAFE experiment \cite{damico2013prisma}. The near and far-field multi-path models assumed are

\begin{equation}
  \label{eq26}
  M_{k,near} = 
  A_k \text{cos}^2 (\varepsilon_q)
\end{equation}

\begin{figure}[h]
	\centering
    \includegraphics[width=0.44\textwidth]{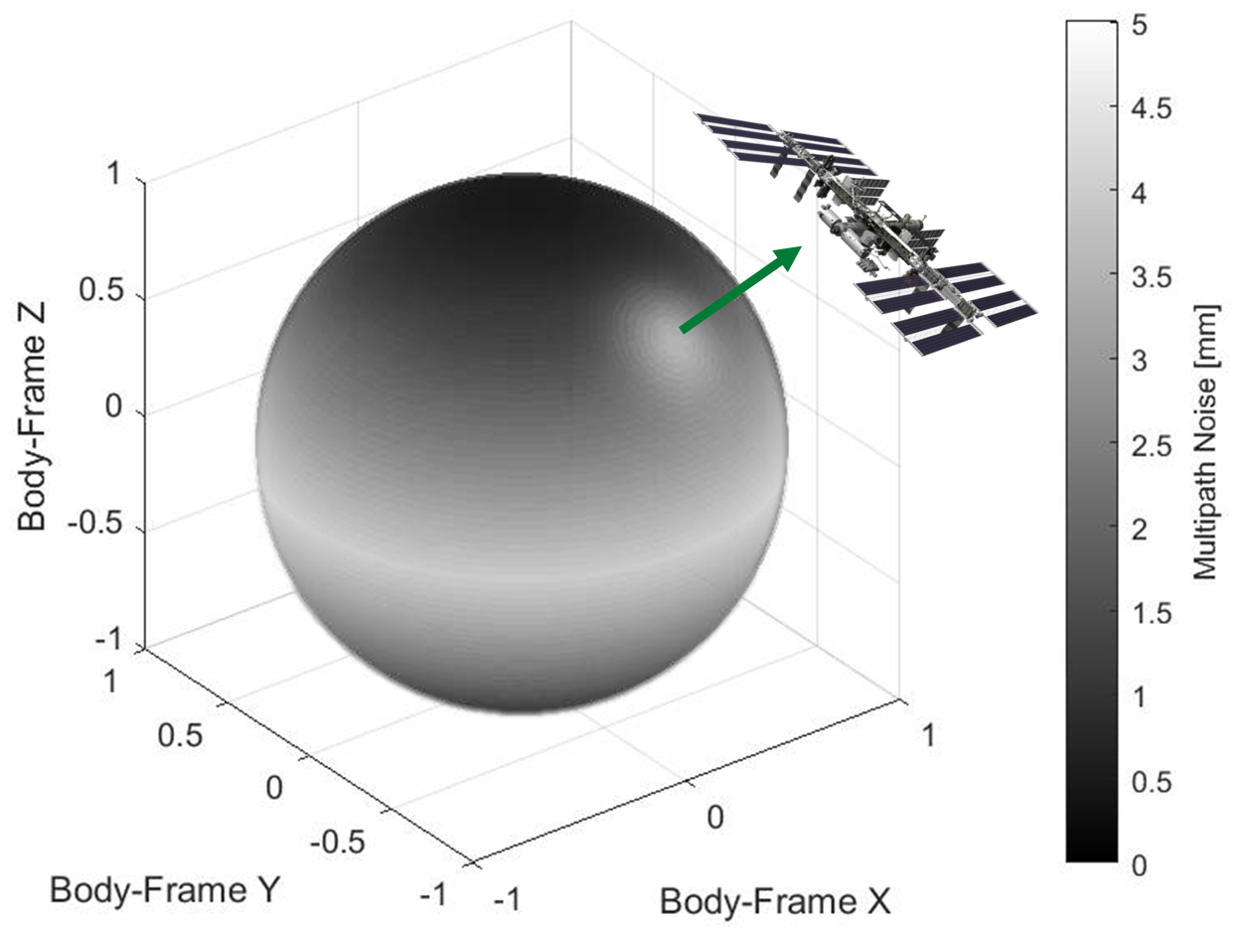}
    \caption{An example multi-path error model on carrier phase, dependent on azimuth-elevation of received signal due to near and far-field (external object) influences.}
    \label{figure:multipath-example}
\end{figure}

\vspace{-5mm}

\begin{equation}
  \label{eq27}
  \footnotesize
  M_{k,far} = 
  \frac{A_k}{(A_k R + 1)^2}
  \exp{ \left[ -\frac{R}{S} \left( 
  (\alpha_q - \alpha)^2 +
  (\varepsilon_q - \varepsilon)^2
  \right) \right]}
\end{equation}

where the net multi-path noise is assumed additive to the measurements; identifier subscripts $k \in [\rho, \phi]$ denotes multi-path influence on code pseudorange ($\rho$) or undifferenced carrier phase ($\phi$); $A_k$ is an amplitude term; $\alpha$ and $\varepsilon$ are the elevation and azimuths of the target spacecraft in the $\upsilon$ frame as per Figure \ref{figure:angles-ref-frame}; $\alpha_q$ and $\varepsilon_q$ are the queried azimuths and elevations corresponding to the direction of arrival of the GPS L1 signal; $S$ is a size factor that scales with the far-field target's physical size, reflecting the size of the multi-path footprint cast by the external body on the receiving antenna’s azimuth-elevation profile as illustrated in Figure \ref{figure:multipath-example}; and $R$ is the range between the local spacecraft to the far-field (target) object. Total multi-path noise on a GPS measurement arriving from $\alpha_q$ and $\varepsilon_q$ directions are computed by sampling a zero-mean Gaussian distribution whose standard deviation is the sum of the individual effects given by equations \ref{eq26} and \ref{eq27}. Parameters $A_k$ and $S$ are selected to match the magnitudes of multi-path effects observed in relevant missions \cite{damico2012safe} \cite{powe2012issmpath} \cite{svehla2018leo}. An example of the multi-path model on carrier phase, projected on the unit attitude sphere, is shown in Figure \ref{figure:multipath-example} with a far-field object (e.g. ISS).

\section{Navigation Performance}
\label{section7}


\subsection{Scenario 1: Rendezvous in LEO}
\label{section6:sc1}

\begin{figure}[h]
	\centering
    \includegraphics[width=0.48\textwidth]{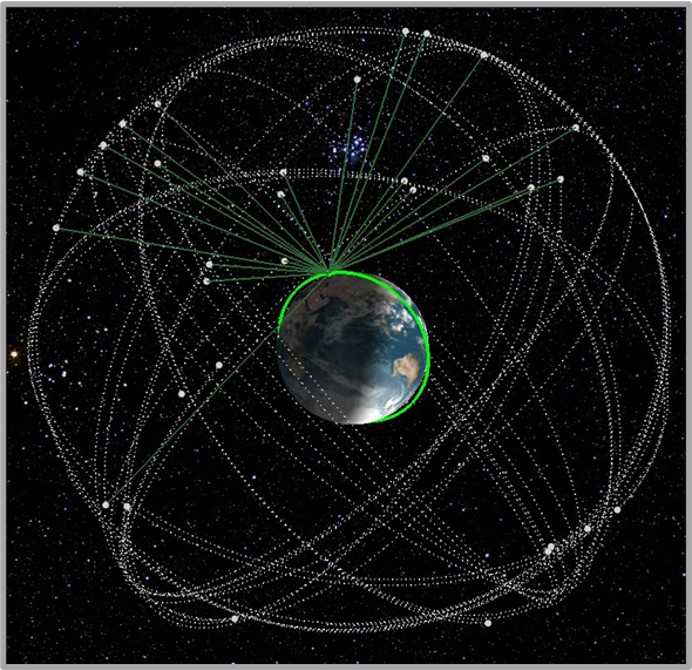}
    \caption{GPS constellation provides good geometry for receivers in LEO but with short time-visibility of tracked integer ambiguities, typically less than 1/3 of an orbit period.}
    \label{figure:flight-sc2-03}
\end{figure}

This scenario is a simulated rendezvous and docking mission in LEO between a chaser spacecraft and the International Space Station (ISS) at a mean altitude of 370km and inclination of $51.6^\circ$. This scenario is characterized by an environment with good geometric diversity in LEO as seen in Figure \ref{figure:flight-sc2-03}, with benign thermal noise on code pseudorange and carrier phase, but afflicted by severe multi-path on both code and carrier due to the large reflective appendages of the ISS, as well as the short time-visibility of tracked integer ambiguities due to the rapidly changing geometry.

GPS L1 measurement updates are applied in the filter at 30s cadence, while external aiding sensors provide measurement updates at 10s cadence. The measurement noise parameters are tabulated in Table \ref{tab:meas-noise-sc2}, where $\sigma_\rho$ and $\sigma_\phi$ are the standard deviations of code pseudorange and undifferenced carrier phase; $S$ is a size factor reflecting the external object's multi-path footprint as per equation \ref{eq27}. 

\begin{table}[h]
\renewcommand{\arraystretch}{1.3}
\caption{\bf Measurement noise parameters in LEO}
\label{tab:meas-noise-sc2}
\centering
\begin{tabular}{|c|c|c|c|c|}
    \hline
    \multicolumn{2}{|c|}{\bfseries Thermal Noise} &
    \multicolumn{3}{|c|}{\bfseries Multipath} \\
    \hline
    $\sigma_\rho$ & $\sigma_\phi$ & $S$ & $A_\rho$ & $A_\phi$  \\
    \hline
    0.20m & 2mm & 5 & 5.0m & 50mm \\
    \hline
\end{tabular}
\end{table}

The GPS measurement thermal noise is derived from the link-budget analysis in Table \ref{tab:link-budget}, while the multi-path parameters were chosen to match similar order-of-magnitude effects observed in measurement residuals of an actual rendezvous mission with the ISS \cite{powe2012issmpath}, in Table \ref{tab:meas-noise-sc2}. Finally, a simplification in the simulation is that bearing-angles are taken with reference to the center of the Zvezda docking port, and thus a key assumption is that bearing angles are resolvable from the full optical image even at close range. 

The initial along-track separation between the chaser and target (ISS) is 1km. A sequence of impulsive control maneuvers are computed open-loop via numerical targeting \cite{berry2011broyden} to bring the chaser to zero terminal relative position and zero contact velocity with respect to the docking port.

\begin{figure}[h]
	\centering
    \includegraphics[width=0.475\textwidth]{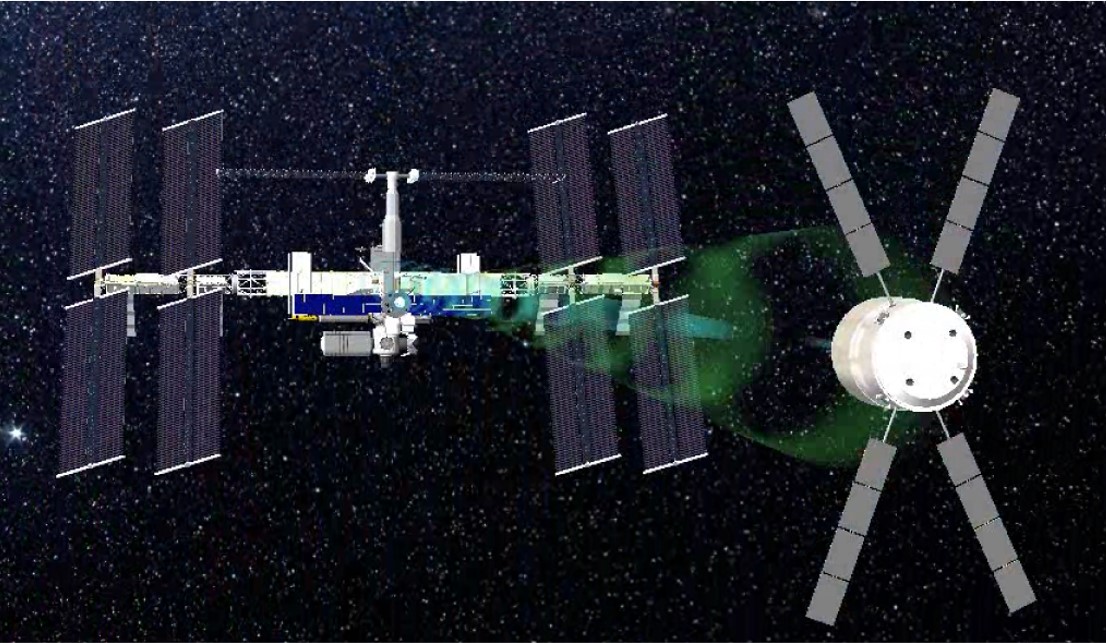}
    \caption{Visualization in STK of the rendezvous operation between local chaser in LEO and target ISS}
    \label{figure:flight-sc2-01}
\end{figure}

\begin{figure}[h]
	\centering
    \includegraphics[width=0.475\textwidth]{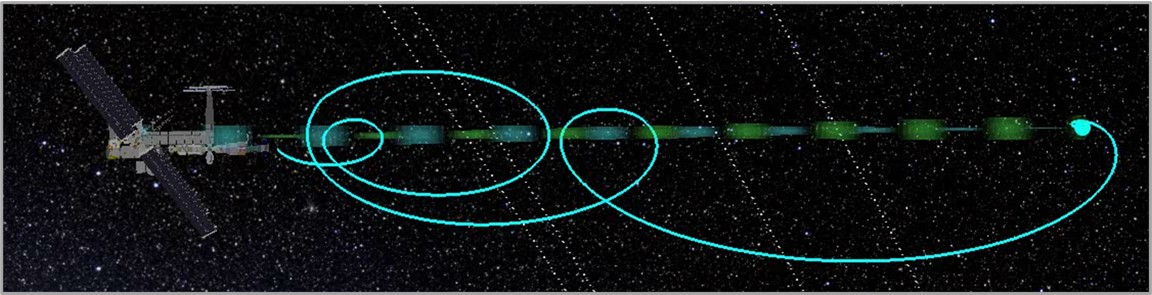}
    \caption{Trajectory of chaser in RTN frame of the ISS, origin centered at the Zvezda docking port}
    \label{figure:flight-sc2-02}
\end{figure}

\begin{figure*}[ht]
    \centering
    \begin{subfigure}[t]{0.47\textwidth}
        \centering
        \includegraphics[width=\textwidth]{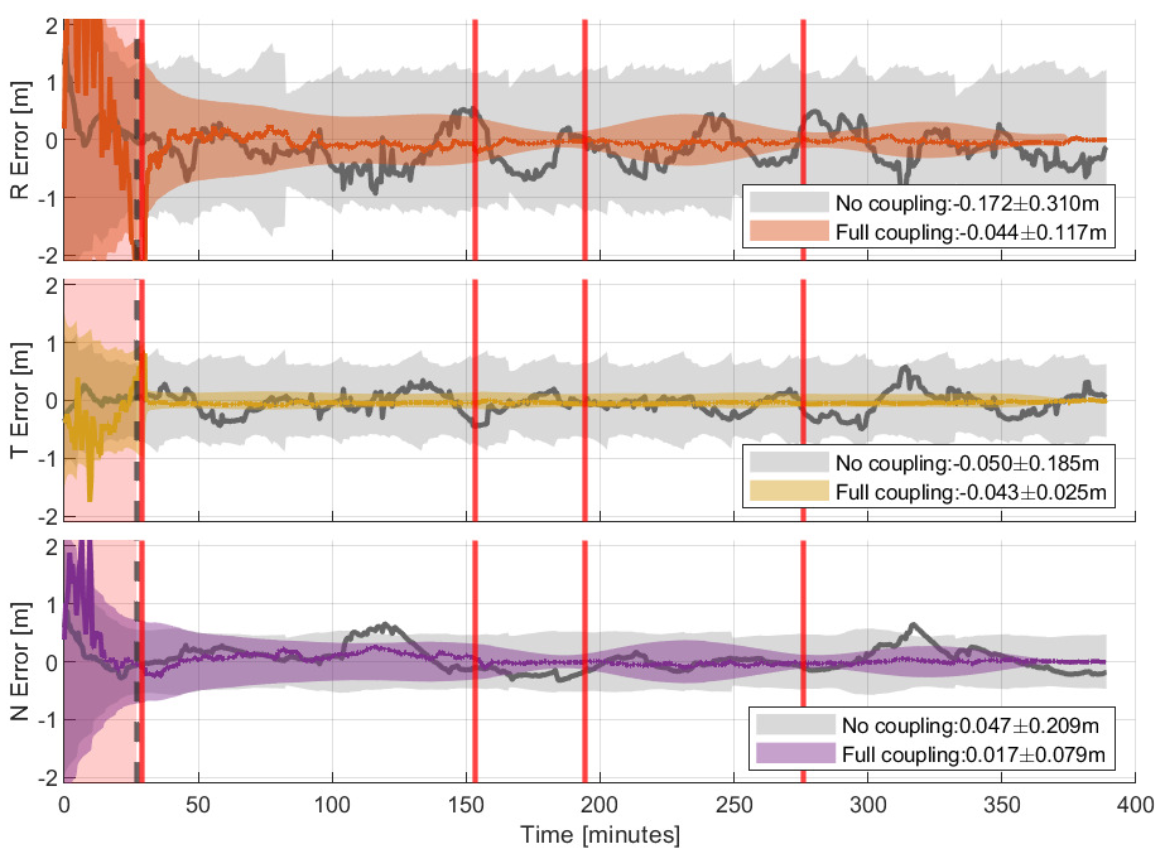}
    \end{subfigure}
    \hfill
    \begin{subfigure}[t]{0.47\textwidth}
        \centering
        \includegraphics[width=\textwidth]{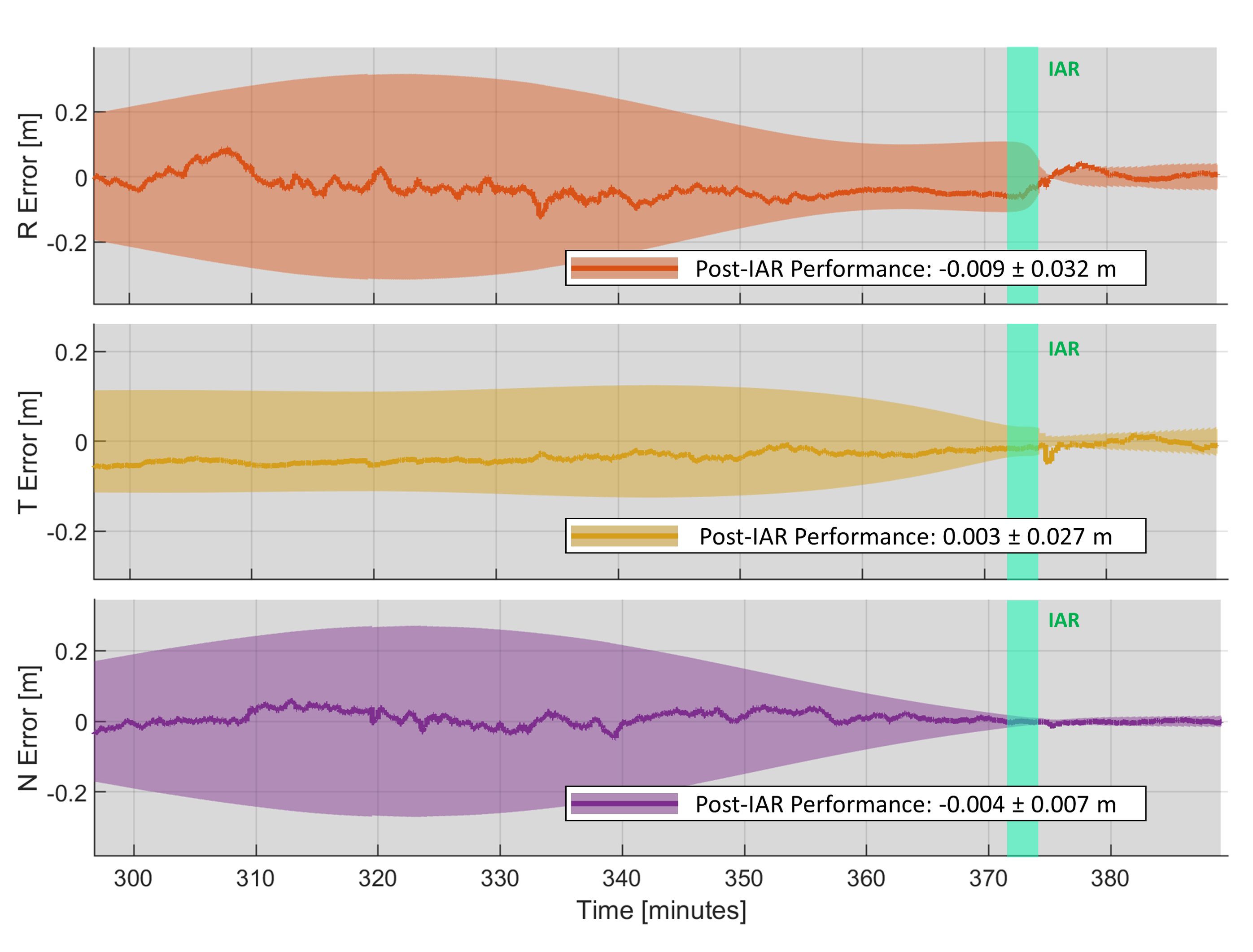}
    \end{subfigure}
    \hfill
    \begin{subfigure}[t]{0.47\textwidth}
        \centering
        \includegraphics[width=\textwidth]{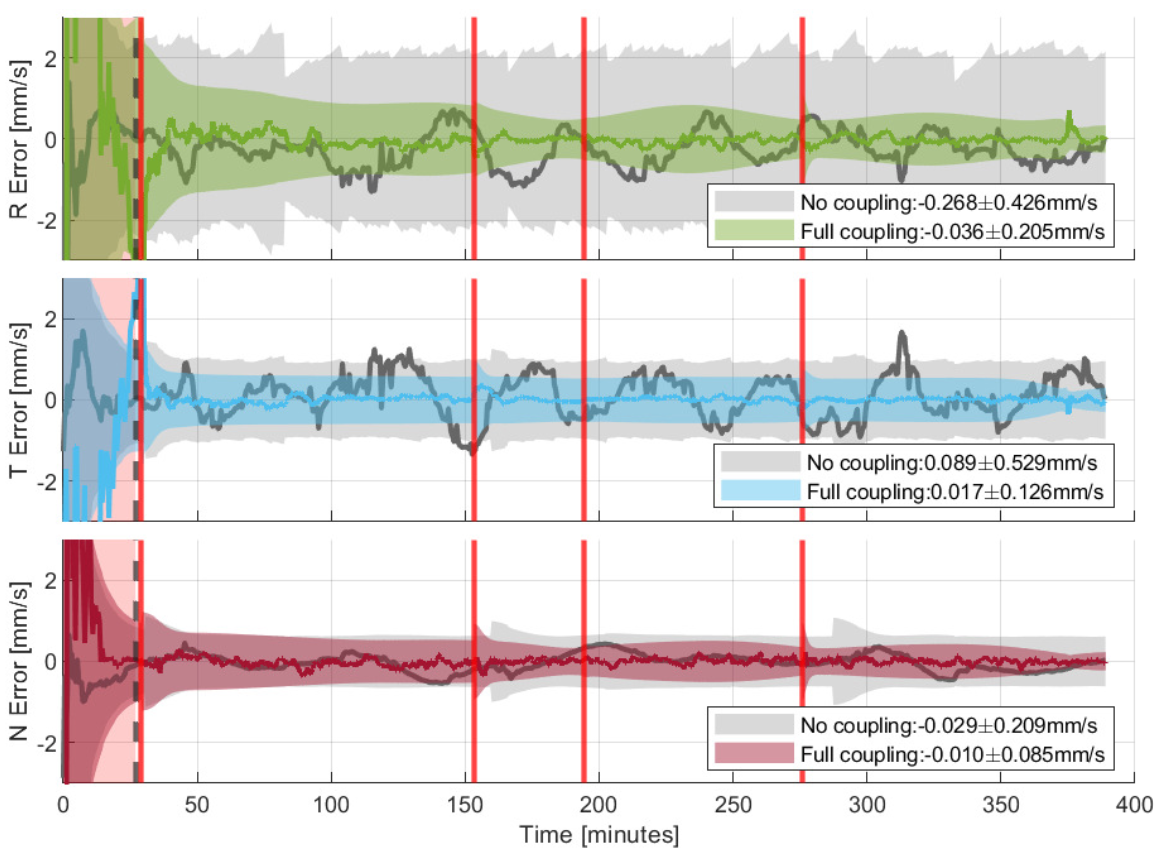}
    \end{subfigure}
    \hfill
    \begin{subfigure}[t]{0.47\textwidth}
        \centering
        \includegraphics[width=\textwidth]{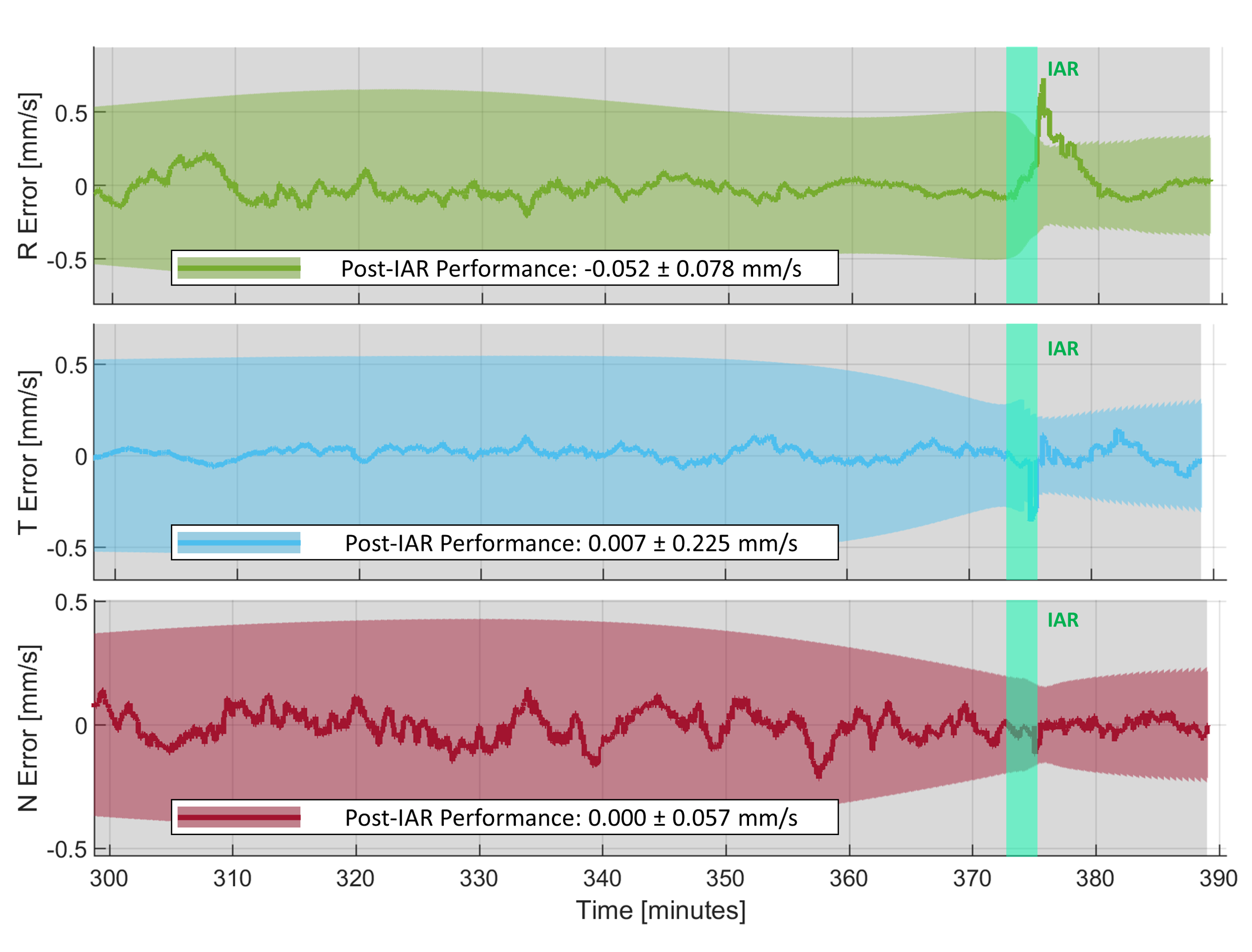}
    \end{subfigure}
    \hfill
    \caption{Navigation performance under under high multi-path during LEO rendezvous and docking: relative position error in RTN throughout the campaign (top-left) and a zoomed-in plot across a 1-orbit period where IAR executes successfully (top-right); relative velocity error in RTN throughout the campaign (bottom-left) and a similar zoomed-in plot (bottom-right). Navigation performance without external sensor coupling is plotted in grayscale while full coupling is plotted in color.}
    \label{figure:results-sc2-rtn}
\end{figure*}

The relative navigation performance in the target's RTN frame is graphed in Figure \ref{figure:results-sc2-rtn}. Maneuvers are indicated by the red vertical bars. Performance under full sensor coupling is given by the colored plots, while performance with no sensor coupling is given by grayscale plots. In the case of full coupling, range and bearing angle measurement updates are made available only 30 minutes into flight. 

\begin{figure*}[ht]
    \centering
    \begin{subfigure}[t]{0.54\textwidth}
        \centering
        \includegraphics[trim={0cm 0.3cm 0cm 0cm}, clip, width=\textwidth]{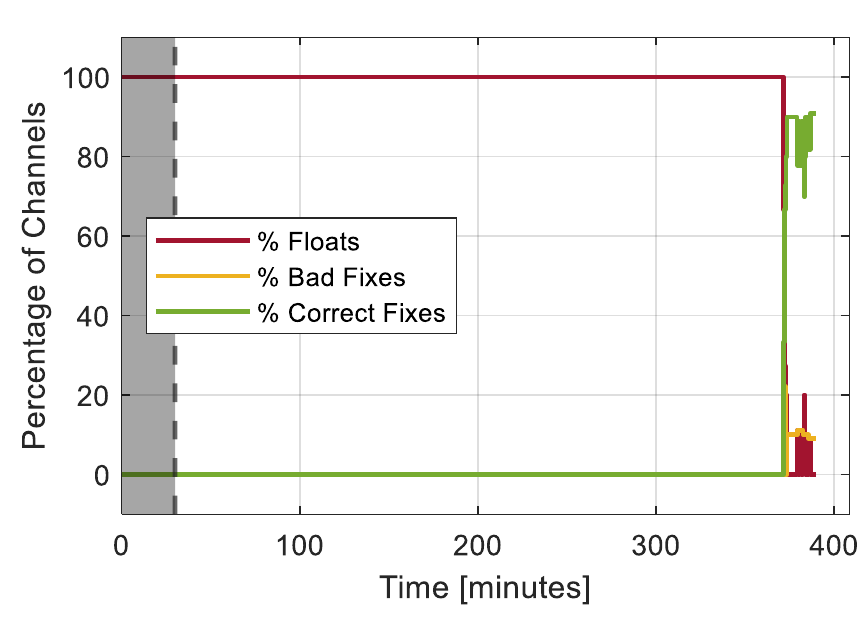}
    \end{subfigure}
    \hfill
    \begin{subfigure}[t]{0.4\textwidth}
        \centering
        \includegraphics[width=\textwidth]{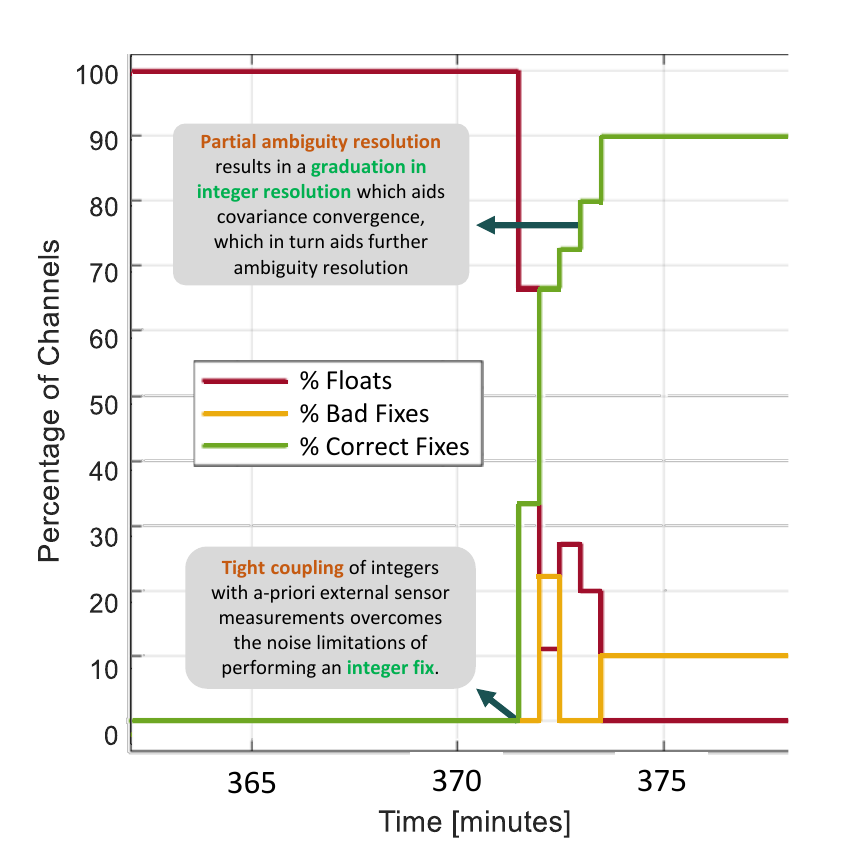}
    \end{subfigure}
    \hfill
    \caption{IAR percentage success (left) under high multi-path during LEO rendezvous and docking, with full sensor coupling, and a zoomed-in plot around the time-at-first-fix at $t = 371$ minutes (right), illustrating the behaviour of graduated integer resolution when resolving ambiguities in partial subsets. Note that at convergence, there is a $10\%$ wrong fix rate.}
    \label{figure:results-sc2-iar}
\end{figure*}

The IAR performance is given in Figure \ref{figure:results-sc2-iar} for only the full-coupling case. As expected, under the prescribed noise conditions in a no-coupling scenario, all SDCP ambiguities remained unresolved and thus plots of IAR performance under no-coupling were excluded. With full coupling, achieving IAR becomes a real possibility due to the presence of measurements not afflicted by high multi-path. This allows (i) the steady-state covariances of all $\Tilde{N}_{SD}$ to converge sufficiently towards passing integer acceptance tests, and (ii) potentially reducing the probability of a wrong integer fix by considering the weighted range and bearing angle information as an a-priori soft constraint. Furthermore, partial resolution rather than full-batch resolution allows for a graduation of integer fixing, aiding filter convergence further towards a successful cascade of resolved integers as depicted in Figure \ref{figure:results-sc2-iar} (right). The time-of-first-fix is $t = 371$ minutes.

\vspace{5mm}

\begin{tcolorbox}
The steady-state relative navigation performance (mean and standard deviation), after IAR, is $1.03 \pm 4.24$ cm for the relative position and $0.0525 \pm 0.245$ mm/s for the relative velocity, root-mean-squared.
\end{tcolorbox}

\clearpage

\clearpage


\subsection{Scenario 2: Rendezvous and Docking in GEO}
\label{section6:sc2}

\begin{figure}[h]
	\centering
    \includegraphics[width=0.48\textwidth]{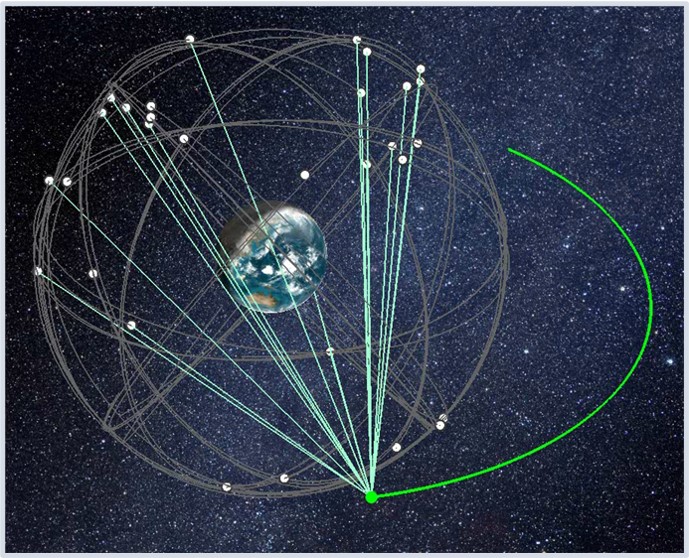}
    \caption{GPS sidelobe-only reception provides poor geometric dilution of precision and low $C/N_0$, but offers a longer window of visibility for tracked ambiguities due to longer orbit periods of the receiver}
    \label{figure:flight-sc3-03}
\end{figure}

This scenario is a simulated rendezvous and docking mission in GEO between a chaser spacecraft and a cooperative micro-satellite with an active inter-satellite cross link. This scenario is characterized by an environment with high thermal noise due to the reliance on GPS sidelobe signals with extremely low $C/N_0$ ratio and poor geometric dilution of precision as seen in Figure \ref{figure:flight-sc3-03}. The maximum half-cone angle of the sidelobe beam is $60^\circ$, with all GPS satellite transmitter boresights pointing in the nadir direction.

It is assumed that the receiver sensitivity can reach $15$ dB-Hz for sidelobe reception \cite{guan2022sidelobegnss}, which is slightly below the expected $C/N_0$ ratio of the GPS L1 sidelobe signal in GEO as tabulated in Table \ref{tab:link-budget}. The cadence of filter measurement updates remains unchanged from the previous scenario. Measurement noise parameters are tabulated in Table \ref{tab:meas-noise-sc3},

\begin{table}[h]
\renewcommand{\arraystretch}{1.3}
\caption{\bf Measurement noise parameters in GEO}
\label{tab:meas-noise-sc3}
\centering
\begin{tabular}{|c|c|c|c|c|}
    \hline
    \multicolumn{2}{|c|}{\bfseries Thermal Noise} &
    \multicolumn{3}{|c|}{\bfseries Multipath} \\
    \hline
    $\sigma_\rho$ & $\sigma_\phi$ & $S$ & $A_\rho$ & $A_\phi$  \\
    \hline
    2.673m & 21.274mm & 1 & 1.0m & 10mm \\
    \hline
\end{tabular}
\end{table}

The GPS measurement thermal noise in Table \ref{tab:meas-noise-sc3} values are derived from the link-budget analysis in Table \ref{tab:link-budget}. Again, it is assumed that bearing-angles are resolvable from the optical image, where angles are taken with reference to the center of the GEO target's docking port. 

\begin{figure}[h]
	\centering
    \includegraphics[width=0.475\textwidth]{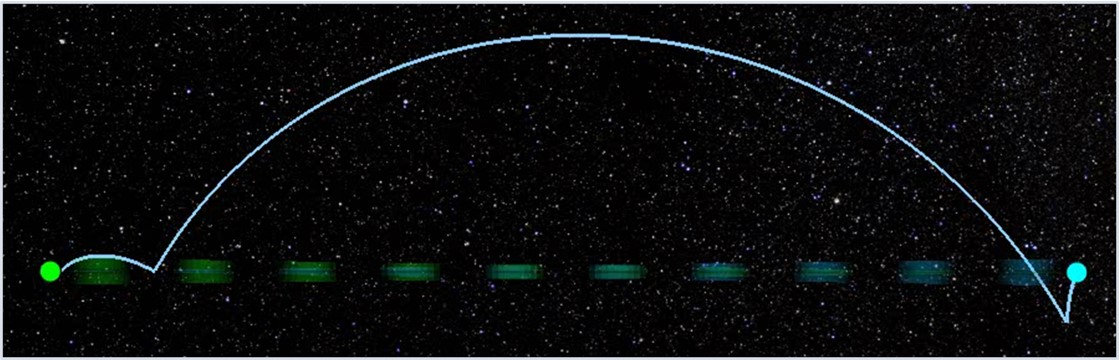}
    \caption{Trajectory of chaser in RTN frame of the GEO target, origin centered at the docking port}
    \label{figure:flight-sc3-02}
\end{figure}

The initial along-track separation between the chaser and target in GEO is 1km. A sequence of impulsive control maneuvers are computed open-loop \cite{berry2011broyden} to bring the chaser to zero terminal relative position and zero contact velocity with respect to the GEO target, resulting in the trajectory seen in Figure \ref{figure:flight-sc3-02}. 

\begin{figure}[h]
	\centering
    \includegraphics[width=0.475\textwidth]{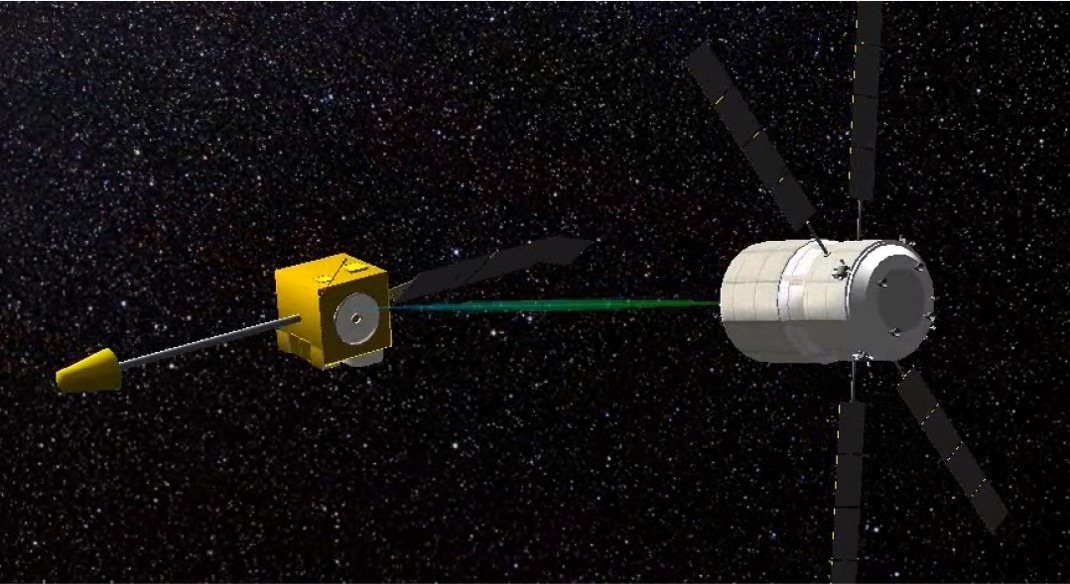}
    \caption{Visualization in STK of the rendezvous operation in geostationary orbit}
    \label{figure:flight-sc3-01}
\end{figure}

\begin{figure*}[hb]
    \centering
    \begin{subfigure}[t]{0.47\textwidth}
        \centering
        \includegraphics[width=\textwidth]{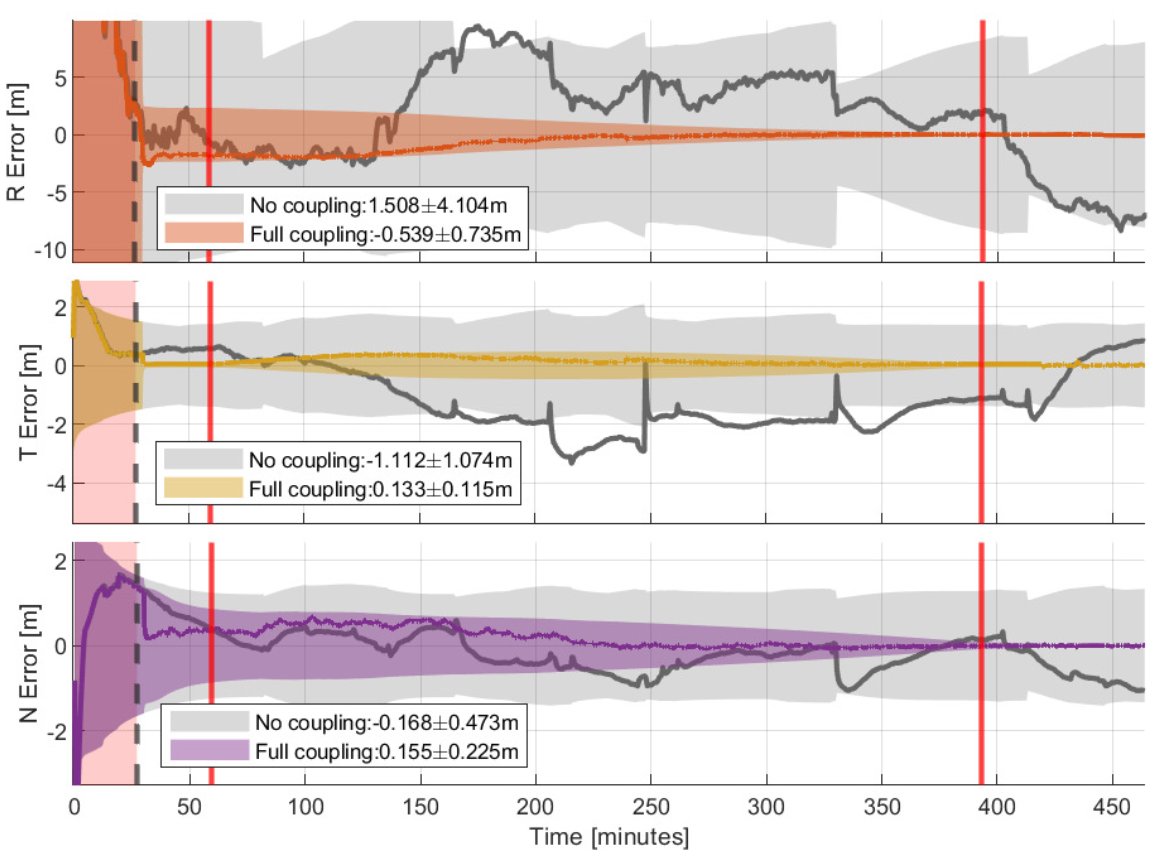}
    \end{subfigure}
    \hfill
    \begin{subfigure}[t]{0.47\textwidth}
        \centering
        \includegraphics[width=\textwidth]{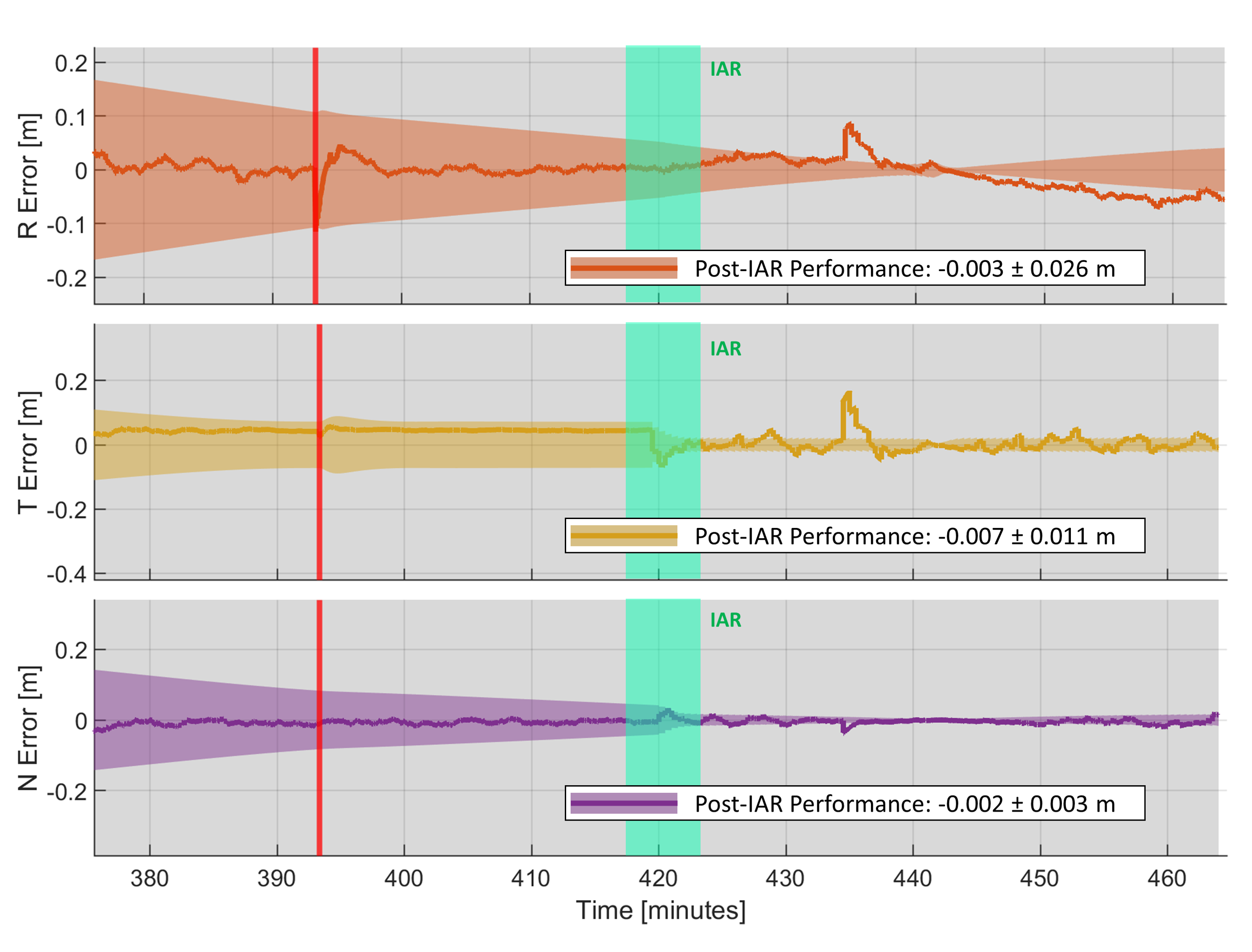}
    \end{subfigure}
    \hfill
    \begin{subfigure}[t]{0.47\textwidth}
        \centering
        \includegraphics[width=\textwidth]{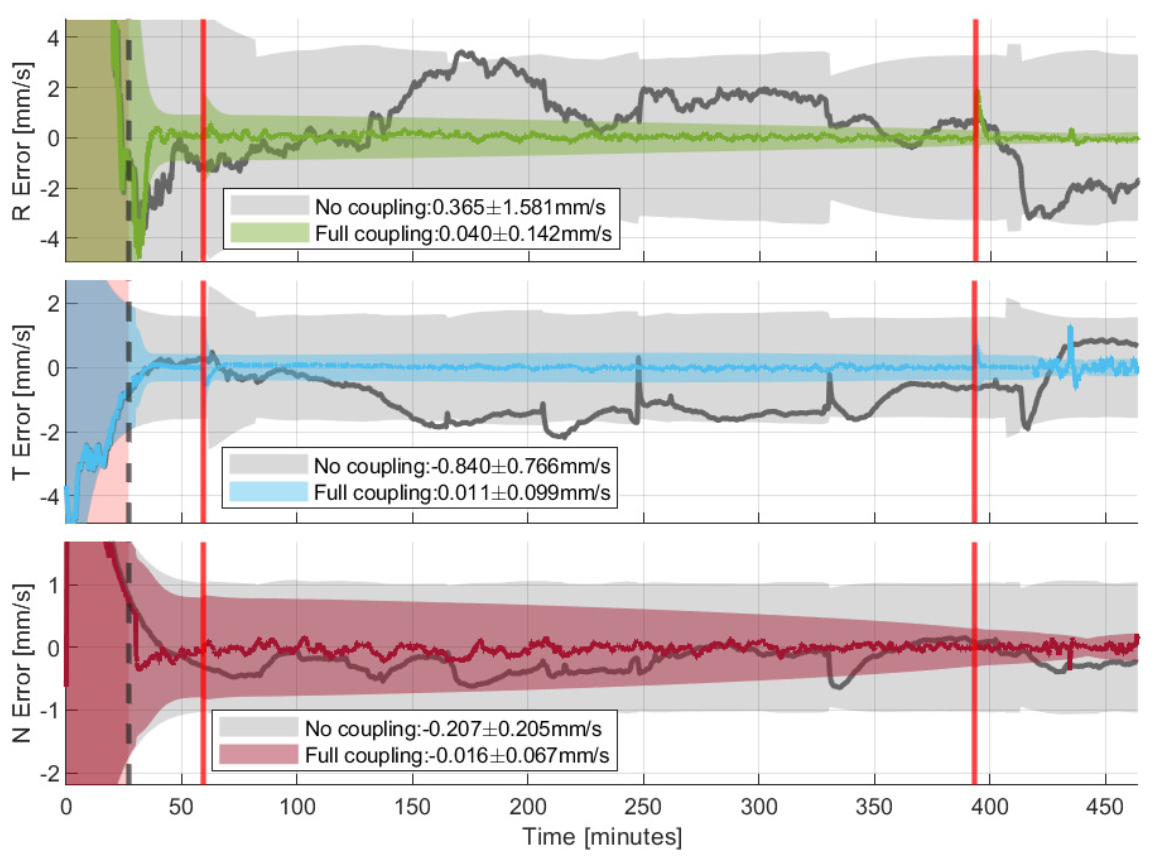}
    \end{subfigure}
    \hfill
    \begin{subfigure}[t]{0.47\textwidth}
        \centering
        \includegraphics[width=\textwidth]{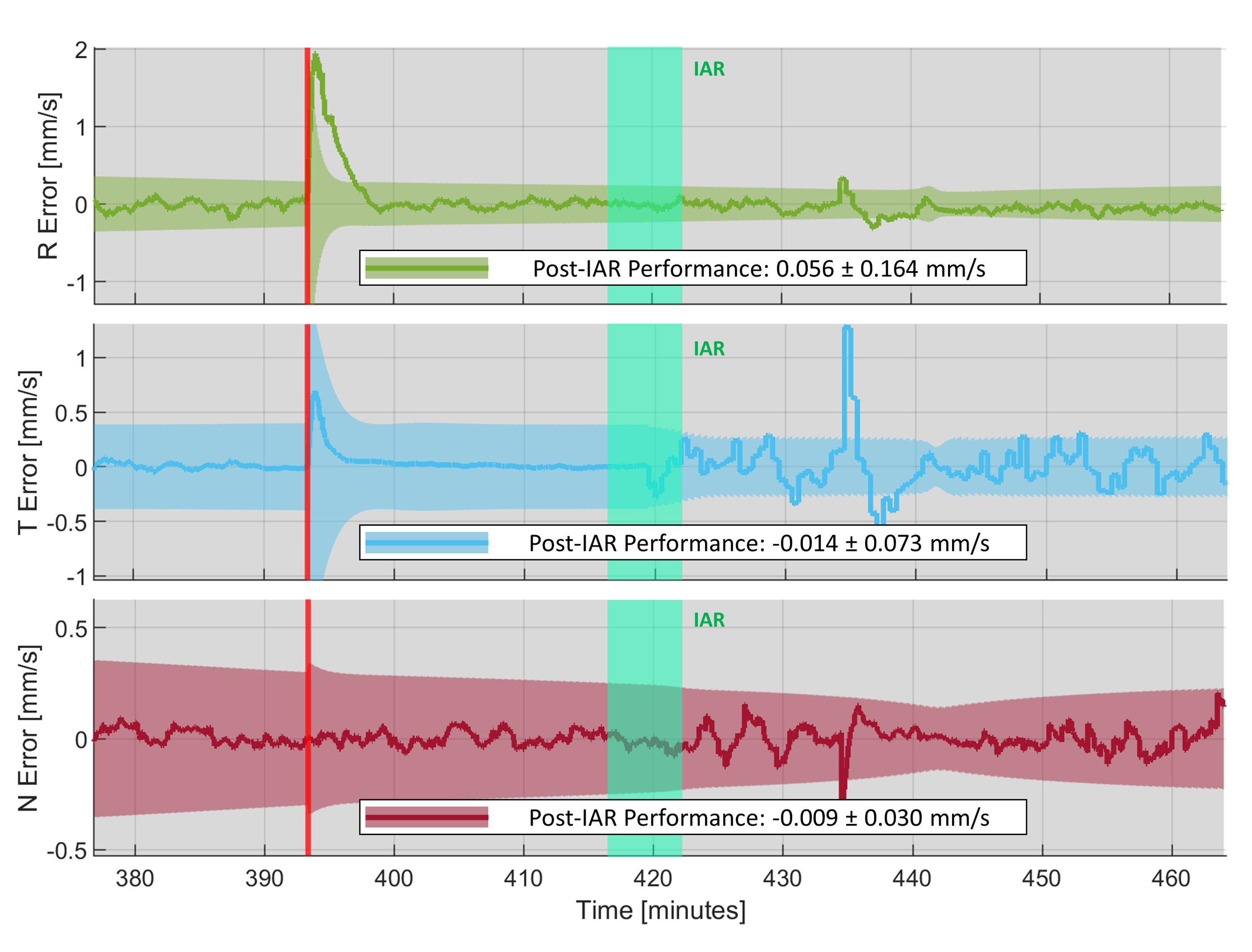}
    \end{subfigure}
    \hfill
    \caption{Navigation performance under high thermal noise of GPS sidelobe-only signals in GEO rendezvous and docking: relative position error in RTN throughout the campaign (top-left) and a zoomed-in plot across a 1-orbit period where IAR executes successfully (top-right); relative velocity error in RTN throughout the campaign (bottom-left) and a similar zoomed-in plot (bottom-right). Navigation performance without external sensor coupling is plotted in grayscale while full coupling is plotted in color.}
    \label{figure:results-sc3-rtn}
\end{figure*}

The relative navigation performance in the target's RTN frame are graphed in Figure \ref{figure:results-sc3-rtn}. Maneuvers are indicated by the red vertical bars. From the zoomed-in plots of Figure \ref{figure:results-sc3-rtn} (right), the maneuvers produce expected and abrupt relative velocity estimation errors. As per the previous scenario, the performance under tight-coupling is given by the colored plots with the colored covariance envelope, while performance with no sensor coupling is given by grayscale plots with grayed covariance envelope. In the case of full coupling, range and bearing angle measurement updates are made available only 30 minutes into flight. A noticeable difference in the navigation performance between the current GEO and the previous LEO scenario, is that the navigation errors in the radial-axis (R) are a lot more pronounced than the along-track (T) and normal (N) axes in the GEO case than the LEO case, in Figures \ref{figure:results-sc3-rtn}. This is expected since from Figure \ref{figure:flight-sc3-03}, the GPS constellation geometry from a GEO orbiter results in a much poorer local vertical dilution of precision (VDOP) under sidelobe-only reception.

\begin{figure*}[hb]
    \centering
    \begin{subfigure}[t]{0.54\textwidth}
        \centering
        \includegraphics[trim={0cm 0.3cm 0cm 0cm}, clip, width=\textwidth]{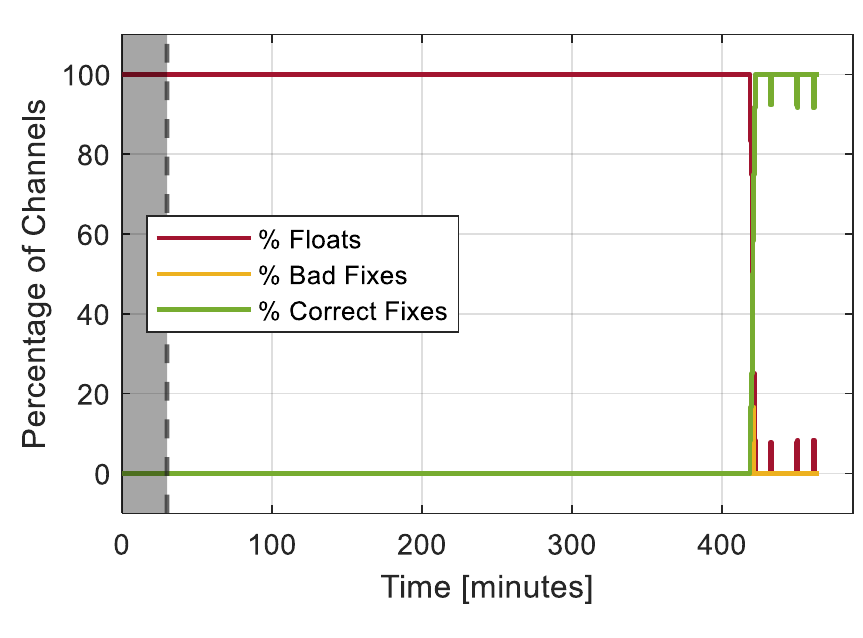}
    \end{subfigure}
    \hfill
    \begin{subfigure}[t]{0.39\textwidth}
        \centering
        \includegraphics[width=\textwidth]{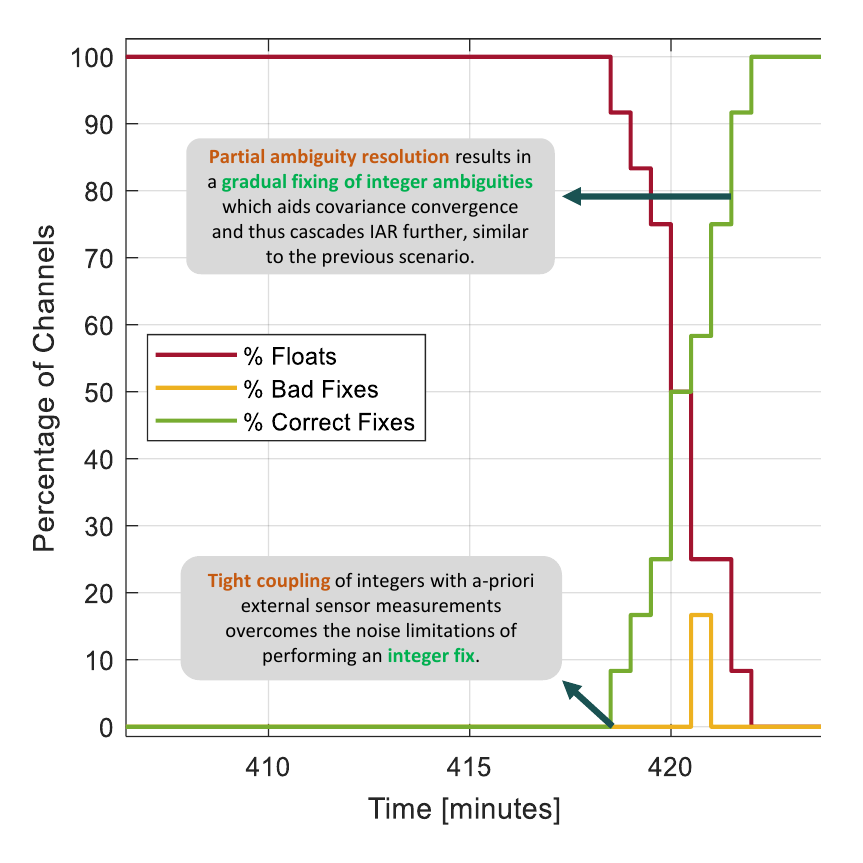}
    \end{subfigure}
    \hfill
    \caption{IAR percentage success (left) under the high thermal noise of GPS sidelobe-only signals in GEO, with full sensor coupling, and a zoomed-in plot around the time-at-first-fix at $t = 418$ minutes (right), illustrating the behaviour of graduated integer resolution when resolving ambiguities in partial subsets.}
    \label{figure:results-sc3-iar}
\end{figure*}

The IAR performance is given in Figure \ref{figure:results-sc3-iar} for only the full-coupling case. Once again, as expected under no-coupling, all SDCP ambiguities remained unresolved as floats and thus plots of IAR performance under no-coupling were excluded. The same advantageous effects of including external aiding measurements in both the loose and tight coupling step are observed in the form of an accelerated IAR time-to-first-fix as well as a graduation of integer resolution, as seen in Figure \ref{figure:results-sc3-iar} (right). The time-of-first-fix in this scenario was $t = 418$ minutes, as seen in right-most plot of Figure \ref{figure:results-sc3-iar}.

\vspace{5mm}

\begin{tcolorbox}
The steady-state relative navigation performance (mean and standard deviation), after IAR, is $0.79 \pm 2.84$ cm for the relative position and $0.0584 \pm 0.182$ mm/s for the relative velocity, root-mean-squared.
\end{tcolorbox}

\clearpage

\clearpage



\section{Conclusion}
\label{section8}

High precision relative navigation between cooperative agents of a Distributed Space System (DSS) can be achieved by leveraging Carrier Phase Differential GPS (CDGPS) measurements with the successful resolution of an ambiguous number of wave cycles within the measurement. This is the well-known Integer Ambiguity Resolution (IAR) problem. Achieving IAR also enables the precise state estimation of other environmental or dynamical parameters of interest. Thus, IAR is crucial for high precision state estimation using GPS/GNSS receivers in general. Yet, attaining IAR under adverse signal noise conditions is challenging due to its high sensitivity to noise.

This work has set out to explore methods for achieving high precision state estimation through IAR while operating in such adverse noise environments, using only GPS L1. The principal innovation in this paper is the design of an integrated three-step navigation architecture, leveraging sensor fusion and coupling between CDGPS measurements and external aiding sensor measurements, sourced from common pre-existing on-board sensors. This architecture was designed with flight-capable software, intended for online and real-time use. A \textit{loose coupling} step fuses all measurements jointly during the measurement update step of an efficiently designed Extended Kalman Filter. A \textit{tight coupling} step directly incorporates information from the external sensors into IAR, by setting a soft constraint on the objective function based on the agreement between the integer candidates and the external sensor measurements. Finally, float ambiguities are resolved in batches of partial subsets for efficiency, in contrast with the more commonly adopted full-batch resolution. These efforts culminated in an integrated navigation architecture that couples sensor measurements with CDGPS, follow by IAR, through a multi-stage process, which is all packaged into a unified and flight-capable software.

The proposed navigation architecture is validated under two case studies with adverse noise conditions in this paper, using state-of-the-art measurement and dynamics modelling with the Stanford Space Rendezvous Lab's $\mathcal{S}$$^3$ library. The first case study is a rendezvous mission in Low Earth Orbit (LEO) with the International Space Station, of which its reflective structures introduces strong multi-path influences on the GPS L1 measurements. The second case study is a rendezvous mission in Geostationary Orbit (GEO) under high thermal noise due to the reliance on GPS sidelobe-only signals at GEO altitudes. High-fidelity simulation demonstrated $< 5$ cm relative position errors and $< 0.25$ mm/s relative velocity errors at filter convergence, with at least $90\%$ IAR success rate. However, achieving IAR under these extremely challenging scenarios requires a long filter convergence time (5 hours for the first case, and 7 hours for the second case) as seen in the flight-like results. Furthermore, wrong integer fixes, which could be detrimental to filter navigation performance, are still a possibility in general, and were observed in the flight results at a $\leq 10\%$ rate. Broadly, the detection of falsely resolved integers remains still an open problem in literature. These gaps therefore motivate potential research directions forward towards both precise \textit{and robust} state estimation in a DSS. Such a direction would also find valued use in many proximity operations scenarios requiring high-precision, such as: autonomous rendezvous and docking, on-orbit maintenance, relative trajectory planning, and improved space domain awareness.


\acknowledgments
The authors of this paper are graciously thankful for the support by the VISORS Mission NSF Award $\#$1936663 and the fellowship support of DSO National Laboratories, Singapore. The authors would also like to acknowledge the Stanford Space Rendezvous Laboratory (SLAB) members for their contributions in developing the flight software ecosystem in SLAB. In particular, sincere appreciation is expressed towards Vincent Giralo and Toby Bell for their contributions in developing the DiGiTaL flight software.


\appendices{}
\label{appendix}

\subsection{External Sensor Measurement Jacobians}

Let $\Vec{\rho}^{(\upsilon)} = [x,y,z]$ be the computed relative baseline vector in the $\upsilon$-frame, and $[R, \alpha, \varepsilon]$ be the range and bearing angle measurements observed, as per Figure \ref{figure:angles-ref-frame}. The Jacobians (sensitivities) of the external aiding sensor measurements with respect to the $\upsilon$-frame coordinates are,

\begin{equation*}
\small
\renewcommand*{\arraystretch}{2}
\mathbf{H}^{\upsilon} =
\begin{bmatrix}
    \frac{\partial R}{\partial x} & 
    \frac{\partial R}{\partial y} & 
    \frac{\partial R}{\partial z} \\
    \frac{\partial \alpha}{\partial x} & 
    \frac{\partial \alpha}{\partial y} & 
    \frac{\partial \alpha}{\partial z} \\
    \frac{\partial \varepsilon}{\partial x} & 
    \frac{\partial \varepsilon}{\partial y} & 
    \frac{\partial \varepsilon}{\partial z} \\
\end{bmatrix}
= \\
\begin{bmatrix}
    \frac{x}{R} &
    \frac{y}{R} &
    \frac{z}{R} \\
    \frac{-xy}{R^3 \cos(\alpha)} &
    \frac{x^2 + z^2}{R^3 \cos(\alpha)} &
    \frac{-yz}{R^3 \cos(\alpha)} \\
    \frac{\cos^2 (\varepsilon)}{z} &
    0 &
    \frac{-x \ \cos^2 (\varepsilon)}{z^2} \\
\end{bmatrix}
\end{equation*}

The Jacobians with respect to the chief and deputy positions in the ECI frame $\Vec{r}_c$ and $\Vec{r}_d$ are expressed as,

\begin{equation*}
\begin{split}
    \mathbf{H}^{ECI}_c & = - \
    \mathbf{H}^{\upsilon} \cdot \mathbf{\Theta}_{ECI}^{\upsilon} \\
    \mathbf{H}^{ECI}_d & = \quad
    \mathbf{H}^{\upsilon} \cdot
    \mathbf{\Theta}_{ECI}^{\upsilon} \\
\end{split}
\end{equation*}

\newpage


\begin{table}[ht!]
\renewcommand{\arraystretch}{1.3}
\caption{\bf Link budget analysis and $C/N_0$ per scenario}
\label{tab:link-budget}
\begin{tabular}{|c|c|c|c|}
    \hline
    \bfseries Parameters & \bfseries Units & \bfseries LEO & \bfseries GEO \\
    \hline\hline
    \footnotesize Frequency
    & MHz & 1575.42 & 1575.42 \\
    \hline
    \footnotesize PLL Bandwidth
    & Hz & 15.0 & 15.0 \\
    \footnotesize RX Antenna Gain
    & dBW & 33.0 & 33.0 \\
    \footnotesize RX Circuit Loss
    & dBW & -1.0 & -1.0 \\ 
    \footnotesize RX Polarization Loss
    & dBW & -1.0 & -1.0 \\
    \hline
    \footnotesize GPS Antenna Gain
    & dBW & 13.5 & -3.0 \\
    \footnotesize GPS Transmit Power
    & dBW & 14.25 & 14.25 \\
    \footnotesize GPS Transmit Loss
    & dBW & -1.25 & -1.25 \\
    \footnotesize GPS EIRP
    & dBW & 26.5 & 10.0 \\
    \hline
    \footnotesize Slant Range
    & km & 20,000 & 80,000 \\
    \footnotesize Free Space Path Loss
    & dBW & -182.419 & -194.460 \\
    \footnotesize Atmospheric Losses
    & dBW & -0.10 & -0.10 \\
    \footnotesize Noise Spectral Density ($N_0$)
    & dBW & -169.919 & -169.919 \\
    \footnotesize Carrier Signal Strength ($C$)
    & dBW & -124.919 & -153.460 \\
    \hline \hline
    \multicolumn{4}{|c|}{\bfseries Resultant Noise} \\
    \hline
    \footnotesize Carrier-to-Noise Ratio $(C/N_0)$
    & dBW & 45.0 & 16.45 \\
    \footnotesize Pseudorange Thermal Noise $\sigma_\rho$
    & m & 0.20 & 2.673 \\
    \footnotesize Carrier Phase Thermal Noise $\sigma_\phi$
    & mm & 2.0 & 21.274 \\
    \hline
\end{tabular}
\end{table}


\begin{table}[ht!]
\renewcommand{\arraystretch}{1.3}
\caption{\bf Navigation Filter Parameters}
\label{tab:filter-params}
\begin{tabular}{|c|c|}
    \hline
    \multicolumn{2}{|c|}{\bfseries Initial Standard Deviation} \\
    \hline
    \footnotesize Position [$m$] & 1000.0 \\
    \footnotesize Velocity [$m/s$] & 1.0 \\
    \footnotesize Emp. Accelerations [$m/s^2$] & $[1.0, 2.0, 0.75] \times 10^{-6}$ \\
    \footnotesize RX clock errors [$m$] & 100.0 \\
    \footnotesize Ambiguities [cycles] & 1000.0 \\
    \footnotesize External Sensor Biases & $10^{-9}$m and $10^{-5}$ arcsec \\
    \hline \hline
    \multicolumn{2}{|c|}{\bfseries Process Noise 1-Sigma} \\
    \hline
    \footnotesize Position [$m$] & $10^{-6}$ \\
    \footnotesize Velocity [$m/s$] & $10^{-9}$ \\
    \footnotesize Emp. Accelerations [$m/s^2$] & $[1.0, 1.0, 0.5] \times 10^{-6}$ \\
    \footnotesize RX Clock Errors [$m$] & 5.0 \\
    \footnotesize Ambiguities [cycles] & 5.0 \\
    \footnotesize External Sensor Biases & $10^{-9}$m and $10^{-5}$ arcsec \\
    \hline \hline
    \multicolumn{2}{|c|}{\bfseries Measurement Noise 1-Sigma} \\
    \hline
    \footnotesize Code Pseudorange [$m$] & 1.5 (LEO), 2.65 (GEO) \\
    \footnotesize Carrier Phase [$mm$] & 15.0 (LEO), 20.0 (GEO) \\
    \footnotesize External Sensors & Assumed known from Table \ref{tab:ext-error-budget} \\
    \hline \hline
    \multicolumn{2}{|c|}{\bfseries Auto-Correlation Time Constants} \\
    \hline
    \footnotesize RX Clock Errors [$s$] & 60.0 \\
    \footnotesize Emp. Accelerations [$s$] & 900.0 \\
    \hline
\end{tabular}
\end{table}

\newpage


\thebiography

\begin{biographywithpic}
{Samuel Y. W. Low}{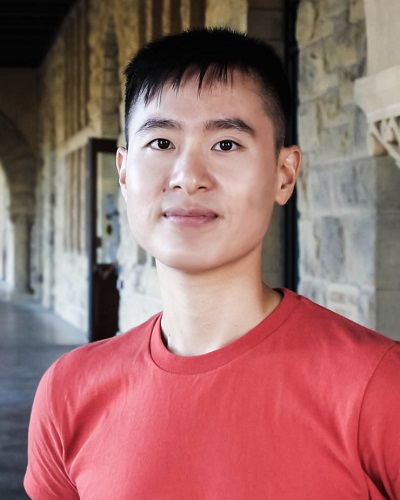} is a Ph.D. Candidate in the Stanford Space Rendezvous Lab (SLAB). He graduated from Stanford with an M.Sc. in Aeronautics and Astronautics (2023) and from the Singapore University of Technology and Design (SUTD) with a B.Sc. in Engineering Product Development (2018). His research focuses on enabling precise and robust state estimation for distributed space systems under realistic and challenging environments using a sensor/data fusion approach, with a focus on solving the Integer Ambiguity Resolution (IAR) problem for Differential GPS. In SLAB, Samuel works on developing the navigation flight software for VISORS, a distributed telescope formation flying mission. Samuel is also a Senior Member of Technical Staff at DSO National Laboratories, Singapore. He had worked on various mission design and analyses trade studies, and also developed precise relative navigation algorithms for Singapore's first formation flying satellite mission. He is a recipient of the DSO Postgraduate Fellowship, the Tan Kah Kee Postgraduate Scholarship, the DSO SOAR Scholarship, and the SUTD Asian Leadership Program Scholarship, jointly awarded by SUTD and Zhejiang University.
\end{biographywithpic} 

\begin{biographywithpic}
{Simone D'Amico}{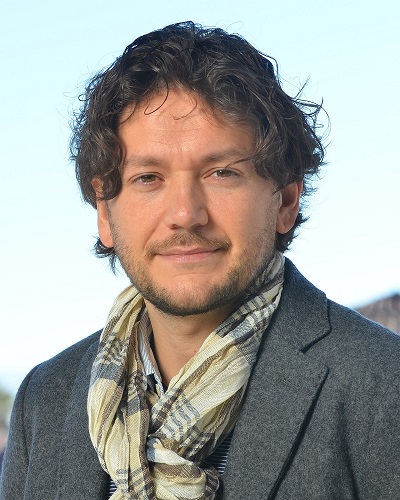} is an Associate Professor of Aeronautics and Astronautics (AA), W.M. Keck Faculty Scholar in the School of Engineering, and Professor of Geophysics (by Courtesy). He is the Founding Director of the Stanford’s Space Rendezvous Laboratory and Director of the AA Undergraduate Program. He received the B.S. and M.S. degrees from Politecnico di Milano (2003) and the Ph.D. degree from Delft University of Technology (2010). Before Stanford, Dr. D’Amico was research scientist and team leader at the German Aerospace Center (DLR) for 11 years. There he gave key contributions to formation-flying and proximity operations missions such as GRACE, PRISMA, TanDEM-X, BIROS and PROBA-3. His research aims at enabling future miniature distributed space systems for unprecedented remote sensing, space and planetary science, exploration and spaceflight sustainability. He performs fundamental and applied research at the intersection of advanced astrodynamics, spacecraft Guidance, Navigation and Control (GNC), autonomy, decision making and space system engineering. Dr. D’Amico is institutional PI of three autonomous satellite swarm missions funded by NASA and NSF, namely STARLING, VISORS, and SWARM-EX. He is Fellow of AAS, Associate Fellow of AIAA, Associate Editor of AIAA JGCD, Advisor of NASA and three space startups (Capella, Infinite Orbits, Reflect Orbital). He was the recipient of several awards, including Best Paper Awards at IAF (2022), IEEE (2021), AIAA (2021), AAS (2019) conferences, the Leonardo 500 Award by the Leonardo da Vinci Society/ISSNAF (2019), FAI/NAA’s Group Diploma of Honor (2018), DLR’s Sabbatical/Forschungssemester (2012) and Wissenschaft Preis (2006), and NASA’s Group Achievement Award for the GRACE mission (2004).
\end{biographywithpic}

\newpage


\bibliographystyle{IEEEtran}
\bibliography{references}

\end{document}